\documentclass[aps,prb,showpacs,amsmath,amssymb,aps,preprintnumbers,twocolumn,superscriptaddress,nofootinbib]{revtex4-2}
\usepackage{amsmath,amssymb}
\usepackage{bm}
\usepackage{tipa}
\usepackage{upgreek}
\usepackage{comment}
\usepackage{mathrsfs}
\usepackage{graphicx}
\usepackage{braket}
\usepackage{enumitem}
\usepackage{mathbbol}
\usepackage{booktabs}
\usepackage{gensymb}
\usepackage{subcaption}
\usepackage[T1]{fontenc}
\usepackage[normalem]{ulem}
\usepackage{color}
\usepackage{xcolor}
\usepackage[colorlinks,bookmarks=true,citecolor=blue,linkcolor=red,urlcolor=blue]{hyperref}
\usepackage{hyperref}

\usepackage{pifont}

\DeclareMathOperator{\tr}{tr}

\begin{document}

\author{Ziwei Wang}
\altaffiliation{Present address: Condensed Matter Theory Center,
Department of Physics, University of Maryland, College Park, Maryland 20742, USA}
\author{Charlie Raca}
\author{Steven H. Simon}

\affiliation{Rudolf Peierls Centre for Theoretical Physics, University of Oxford, Oxford OX1 3PU, United Kingdom}

		\date{\today}

\title{Symmetry and Quantum Geometry in Bloch Bands}

\begin{abstract}

Quantum geometric quantities have featured heavily in the discussion of the properties of quantum systems in recent years. Among quantities most commonly discussed is the variance of Berry curvature and the integral of the trace of the quantum metric tensor. Despite their usefulness, it is known that they suffer from one significant complication: for a tight-binding model, the quantum geometric quantities depend not only on the parameters of the tight-binding model itself, but also the real space geometry of the tight-binding model, the so-called ``orbital embedding''. One explicitly geometry-independent quantity is therefore the minimal value of the quantum geometric quantity out of all possible real space geometries. In this work, we demonstrate that, if the tight-binding model is compatible with certain spatial symmetries, then the real space geometry that minimizes the variance of the Berry curvature or the integral of the trace of the quantum metric tensor must obey all those spatial symmetries. We further show that the statement is applicable to systems with magnetic translation symmetries and other composite symmetries, with implications for the quantum geometry of the Hofstadter model.

\end{abstract}

\maketitle

\section{Introduction}

A Bloch band is characterized by its dispersion $\epsilon(\bm{k})$ and its Bloch wavefunctions $\ket{\psi(\bm{k})}$. While the earliest studies in condensed matter physics were focused on $\epsilon(\bm{k})$, it has long been realized that the Bloch wavefunctions $\ket{\psi(\bm{k})}$, whose properties may be captured by quantum geometric measures, also contain crucial information about the system.  One of the most well-known quantum geometric measures is the Berry curvature, and the relation between the integral of the Berry curvature over the Brillouin zone and the Hall response of an two-dimensional insulator~\cite{thouless_quantized_1982} is one of the most celebrated results in condensed matter physics. By now, it is well-known that quantum geometry plays a significant role in a wide range of phenomena, such as semi-classical transport, electric polarization, and orbital magnetism (see Refs.~\cite{xiao_berry_2010,vanderbilt_berry_2018} for comprehensive reviews). More closely related to this work, the fluctuation of Berry curvature and the ``integrated quantum metric'' (this term is a shorthand for the integral of the trace of the quantum metric tensor; see the definition in Section~\ref{sec:summary}) have been conjectured to determine the ability of a Chern band to stabilize a fractional Chern insulator~\cite{parameswaran_fractional_2012, goerbig_fractional_2012, wu_zoology_2012, shankar_murthy, dobardzic_geometrical_2013, claassen_position-momentum_2015, jackson_geometric_2015, Bauer_2016, lee_band_2017,ozawa_relations_2021, mera_uniqueness_2024,shi_effects_2026}. Furthermore, Chern bands that saturate a certain lower bound on its integrated quantum metric are termed ``ideal bands'', whose properties often mirror those of Landau levels~\cite{roy_band_2014, wang_exact_2021, mera_kahler_2021} (also see its applications in moir\'e materials~\cite{wang_hierarchy_2022,ledwith_family_2022} and other related works~\cite{mera_engineering_2021,ledwith_vortexability_2023,wang_origin_2023, liu_theory_2025,fujimoto_higher_2025}). The integrated quantum metric is also featured within other contexts, such as the superfluid weight of a flatband superconductor~\cite{peotta_superfluidity_2015, julku_geometric_2016, liang_band_2017, herzog-arbeitman_superfluid_2022,huhtinen_revisiting_2022,tam_geometry-independent_2024,herzog-arbeitmanManybodySuperconductivityTopological2026} and the gauge-invariant component of the spread of a Wannier function~\cite{marzari_maximally_1997}.

A tight-binding model, which is specified by a set of hopping amplitudes between orbitals, is frequently the starting point of a calculation in condensed matter physics. However, it was realized that, for a tight-binding model, the quantum metric tensor and Berry curvature are dependent on the orbital positions of the model~\cite{fruchart_parallel_2014, dobardzic_effective_2014, lim_geometry_2015,jackson_geometric_2015, cooper_topological_2019, simon_contrasting_2020} (also see an analogous issue in one-dimensional insulators~\cite{fuchs_orbital_2021}). Certain physical quantities may have explicit dependence on the orbital positions, in which case the appropriate choice is naturally the physical orbital positions (see the discussion in Ref.~\cite{simon_contrasting_2020}). However, for many other situations, one is interested in the properties of the model, such as the ability to stabilize fractional Chern insulators, that should not depend on the real-space geometry. As many tight-binding models have a ``natural'' choice of orbital positions (for example, the Haldane model~\cite{haldane_model_1988} is placed on a honeycomb lattice), which is also typically the most symmetric one, early literature has mostly implicitly assumed such a choice when computing quantum geometric properties. A more critical examination is carried out in Ref.~\cite{huhtinen_revisiting_2022,herzog-arbeitmanManybodySuperconductivityTopological2026}, where the authors showed that, for the purpose of computing the geometric contribution to the superfluid weight under the mean-field limit, the correct orbital positions are the ones that minimize the integrated quantum metric. The authors also showed that for bands that satisfy the so-called ``uniform pairing condition''\footnote{Let $P_{ab}(\bm{k})$ be the single-particle density matrix that corresponds to fully occupying the relevant bands, where $a,b$ are atomic basis indices. The ``uniform pairing condition'' means that $\int_{\text{BZ}} d^2 \bm{k} \ P_{aa}(\bm{k})$ is an $a$-independent constant.}, orbital positions should preserve the spatial symmetries of the tight-binding model in order to minimize the integrated quantum metric.

In this work, we show that, \textit{to minimize the variance of Berry curvature or the integrated quantum metric for a given tight-binding model, orbital positions should be chosen to obey the symmetries that the tight-binding model is compatible with} (see the exact formulation in Section~\ref{sec:summary}). For the case of the integrated quantum metric, our work recovers the aforementioned result in Ref.~\cite{huhtinen_revisiting_2022,herzog-arbeitmanManybodySuperconductivityTopological2026} as a special case, with our result being also applicable to bands that do not satisfy the ``uniform pairing condition''. We also show that, after excluding certain trivial exceptions, the set of orbital positions that minimize the integrated quantum metric is \textit{unique} up to an overall translation, answering the question raised in Ref.~\cite{tam_geometry-independent_2024}. Furthermore, while almost all previous works are exclusively concerned with the positions of the orbitals within the unit cell, in this work we also consider and answer the more general problem of optimizing quantum geometric measures over \emph{all} lattice geometry, including both the shape of the unit cell and positions of orbitals within it.

We then extend the discussion to include gauge transformations. Many models obey symmetries that are compositions of spatial transformations and gauge transformations, and we show that our discussion is equally applicable to such cases. For a Hofstadter model, the ``natural'' choice to place the orbitals within a magnetic unit cell is to place them on a regular grid so that magnetic translation symmetries are manifestly obeyed. One of the implications of our results is that this ``natural'' choice actually minimizes the variance of Berry curvature and the integrated quantum metric.

Our discussion can be readily applied to the study of ideal bands. An ideal band, as mentioned, is one whose integrated quantum metric saturates a certain lower bound that is determined solely by the Chern number. This naturally implies that out of all possible choices of orbital positions, the choice for which the model is ideal must minimize the integrated quantum metric. Following the preceding discussion, we therefore conclude that an ideal band's orbital positions must obey the symmetry that the tight-binding model is compatible with. We illustrate this point with the Kapit-Mueller model~\cite{kapit_exact_2010}.

In Section~\ref{sec:summary}, we set up the problem carefully and state our main conclusion. In Section~\ref{sec:proof}, we prove the basic version of our statement (without considering gauge transformations) and unpack its implications in Section~\ref{sec:Wyckoff}. In Section~\ref{sec:gauge}, we extend our discussion to include gauge transformations. In Section~\ref{sec:corollaries}, we discuss the application of our results to ideal bands and some other situations. In Section~\ref{sec:conclusion}, we present the conclusion.

\section{Summary of results}\label{sec:summary}

\subsection{Setup}\label{subsec:setup}

A tight-binding model is specified by a set of atomic orbitals connected by some hopping amplitudes. We denote the atomic orbitals by Greek letters $\alpha, \beta$ etc., and the corresponding hopping amplitudes by $t_{\alpha,\beta}$, which obey $t_{\beta,\alpha} = t^*_{\alpha, \beta}$. Since the tight-binding model describes a crystalline system, it is possible to write the orbital index as a tuple of indices $(m,n,a)$, where $m ,n$ take on arbitrary integer values and can be interpreted as the unit cell labels, and $a$ takes on a finite range of values and can be interpreted as the labels of orbitals inside a unit cell (here we have assumed that the system is two-dimensional, though much of the discussion in this work can be generalized to higher dimensions). The hopping amplitudes can then be written as $t_{(m,n,a), (m',n',a')}$. The periodic nature of the system implies
\begin{equation}
    t_{(m,n,a), (m',n',a')} = t_{(m-m',n-n',a), (0,0,a')}.
\end{equation}
The way to decompose the atomic orbital label $\alpha$ into $(m,n,a)$ is not unique, as the choice of unit cell is not unique. Nevertheless, while the results are independent of this choice, it is frequently convenient to make a definite choice.

A real-space realization (also called an ``embedding'' for brevity) of the tight-binding model consists of assigning a particular position $\bm{r}_\alpha$ to each atomic orbital $\alpha$. For our purposes, we will usually only consider real-space realizations that obey translation symmetries, i.e., for some unit cell convention, we specify $\bm{a}_1$, $\bm{a}_2$ and $\bm{r}_a$ such that the orbital positions are given by
\begin{equation}\label{eq:aar}
    \bm{r}_{m,n,a} = m\bm{a}_1 + n\bm{a}_2 + \bm{r}_a.
\end{equation}

Consider a spatial operation $g$ whose action on the real space is denoted as $\bm{r} \rightarrow g(\bm{r})$. The real-space model obeys a spatial symmetry $g$ if and only if there exists some permutation on the set of orbitals $\sigma$ such that 
\begin{align}
    \bm{r}_{\sigma(\alpha)} &= g(\bm{r}_\alpha), \label{eq:positions}\\
    t_{\sigma(\alpha), \sigma(\beta)} &= t_{\alpha, \beta} \label{eq:hopping}.
\end{align}
We emphasize that the spatial symmetry of a real-space model depends not only on the positions $\bm{r}_a$ but also the hopping amplitudes $t_{\alpha,\beta}$, and when we say that a set of orbital positions is \textit{symmetric}, it is always understood with respect to some set of hopping amplitudes. We say that a tight-binding model is \textit{compatible} with a set of spatial symmetries if and only if there exists some embedding $\bm{r}_\alpha$ such that the corresponding real-space model obeys the set of spatial symmetries in the sense defined above.

In the practical study of condensed matter physics, one frequently encounters systems which exhibit symmetries that are compositions of spatial operations and gauge transformations. Here, by ``gauge transformation'' we specifically mean a real-space gauge transformation of the form
\begin{equation}\label{eq:gauge}
    \hat{c}_{\alpha} \rightarrow \hat{c}_{\alpha}e^{i\phi_\alpha}.
\end{equation}
Then, a real-space model is said to obey a spatial symmetry \textit{up to gauge} if Eq.~\ref{eq:hopping} is replaced with
\begin{equation}\label{eq:hopping_gauge}
    t_{\sigma(\alpha), \sigma(\beta)} = t_{\alpha, \beta}e^{-i(\phi_\alpha - \phi_\beta)}.
\end{equation}
An example of this is the Hofstadter model, where a translation combined with a gauge transformation is a symmetry. Similarly, a real-space model obeys a spatial symmetry up to gauge and time-reversal if Eq.~\ref{eq:hopping} is replaced with
\begin{equation}\label{eq:hopping_trs}
    t_{\sigma(\alpha), \sigma(\beta)} = t^*_{\alpha, \beta}e^{-i(\phi_\alpha - \phi_\beta)},
\end{equation}
instead. Here, we have assumed a spinless system such that time-reversal symmetry is simply the complex conjugation operator under the real space basis. A tight-binding model is said to be compatible with a spatial symmetry up to gauge (and time-reversal) if there exists a real-space realization that obeys the spatial symmetry up to gauge (and time-reversal).

\subsection{Main statement}

We are interested in the behaviour of quantum geometric properties of a Bloch band\footnote{The discussion can be easily extended to the case of a group of Bloch bands. For simplicity, we mostly keep to the notation of a single band in the main text.} under change of real-space realization of the same tight-binding model. First off, we define the quantities of interest. The quantum geometric tensor is defined as
\begin{equation}
    Q_{ij} = \braket{\partial_{k_i}u(\bm{k})|Q(\bm{k})|\partial_{k_j}u(\bm{k})},
\end{equation}
where $Q(\bm{k}) = 1 - \ket{u(\bm{k})}\bra{u(\bm{k})}$. Using this, we can define other useful quantities. More specifically, the quantum metric tensor is the real-part of the quantum geometric tensor, i.e.
\begin{equation}
    g_{ij} = \Re[Q_{ij}],
\end{equation}
while the Berry curvature is given by the imaginary part of its off-diagonal entry
\begin{equation}
    \Omega = -2\Im[Q_{xy}]
\end{equation}

Frequently, it is useful to consider quantities integrated over the Brillouin zone (BZ). In this work, we primarily consider the following two quantities: the integral of the trace of the quantum metric tensor, which will be called the ``integrated quantum metric'' for brevity, given by
\begin{equation}
    I \equiv \int_{\text{BZ}}d^2\bm{k} \ \tr[g],
\end{equation}
and the (dimensionless) variance of Berry curvature
\begin{equation}
    \sigma^2_{\Omega} \equiv \frac{A_{\text{BZ}}}{4\pi^2}\int_{\text{BZ}}d^2\bm{k} \ (\Omega(\bm{k}) - \bar{\Omega})^2,
\end{equation}
where $A_{\text{BZ}}$ is the area of the Brillouin Zone and $\bar{\Omega}$ is the average Berry curvature $\bar \Omega = 2 \pi C / A_{\text{BZ}}$ with $C$ the Chern number.

The central claim of this work is that \textit{in order to minimize the integrated quantum metric or the variance of Berry curvature, the real-space realization of a tight-binding model must obey all the spatial symmetries that the tight-binding model is compatible with up to gauge transformation and possible time-reversal operation}\footnote{In the fine-tuned cases where there are more than one global minima (after accounting for overall translations), at least one of the global minimum must obey the spatial symmetry. See the discussion at the end of Section~\ref{sec:proof} and an example in Appendix~\ref{app:flat_B}.}. In the following sections, we will unpack the statement, prove it, and discuss its implications.

\section{Outline of the Proof}\label{sec:proof}

In this section, we outline the proof of the statement made in the preceding section, but restricting to the case of pure spatial symmetry (i.e. one that does not involve gauge transformation or time-reversal). The issue of combined symmetries will be discussed in Section~\ref{sec:gauge}.

The real-space realization of a tight-binding model involves both choosing the lattice vectors $\bm{a}_i$ and the positions of orbitals within the unit cell $\bm{r}_a$ (see Eq.~\ref{eq:aar}). For later reference, we sometimes write $\bm{r}_a = \tilde{x}_{a}\bm{a}_1 + \tilde{y}_a\bm{a}_2$, and call $\tilde{x}_{a}$ and $\tilde{y}_{a}$ the \textit{relative} embedding. One can easily show that the variance of Berry curvature only depends on the relative embedding but not $\bm{a}_i$. However, for the integrated quantum metric $I$, $\bm{a}_i$ can have a non-trivial effect. We will defer a discussion of this issue to the end of the section. Here, we will assume that $\bm{a}_i$ have been chosen to be compatible with the symmetry of the tight-binding model (in more technical wordings, $\bm{a}_i$ have been chosen such that it is possible to further choose appropriate $\bm{r}_a$ such that $\bm{r}_{m,n,a}$ as defined in Eq.~\ref{eq:aar} are symmetric in the sense of Eq.~\ref{eq:positions} and Eq.~\ref{eq:hopping}).

In order to see how the quantum geometric tensor transforms under a change of intra-unit-cell orbital positions, we only need to know how the wavefunctions $u_a(\bm{k})$ transform. This is given by

\begin{equation}\label{eqn:transform}
    \tilde{u}_a(\bm{k}) = e^{-i \bm{k} \cdot \bm{r}_a}u_a(\bm{k}),
\end{equation} and its derivative transforms as
\begin{equation}
    \partial_{k_j}\tilde{u}_a(\bm{k}) =  e^{-i \bm{k} \cdot \bm{r}_a} (\partial_{k_j}u_a(\bm{k})  - ir_{a,j} u_a(\bm{k})).
\end{equation}
Therefore, the quantum geometric tensor transforms as
\begin{widetext}
    \begin{equation}
        \tilde{Q}_{ij}(\bm{k}) = \sum_a \sum_b (\partial_{k_i}u^*_a(\bm{k}) + ir_{a,i}u_a^*(\bm{k}))(\delta_{ab} - u_a(\bm{k})u^*_b(\bm{k}))(\partial_{k_j}u_b(\bm{k}) -ir_{b,j}u_b(\bm{k})).
    \end{equation}
\end{widetext}

From this, we see that the quantum metric tensor depends on the orbital positions $r_{a,i}$ as a polynomial up to the second order. It is also straightforward to show that Berry curvature depends on the orbital positions only up to linear order,
\begin{equation}
    \tilde{\Omega}(\bm{k}) = \Omega(\bm{k}) - \sum_a \epsilon^{ij}r_{a,i}\partial_{k_j}|u_a(\bm{k})|^2.
\end{equation}
This implies that both $I$ and $\sigma^2_{\Omega}$ are polynomial of orbital positions up to the second order. Since both quantities are bounded below, the quadratic part of the polynomial in $r_{a,i}$  must be positive semi-definite. Therefore, there must be a global minimum of either $I$ or $\sigma^2_{\Omega}$ in the space of all embeddings, up to the addition of a null-vector of the quadratic part of the polynomial\footnote{If a vector is a null-vector of the quadratic part, it must also be a null-vector of the linear part, since the total quantity is bounded-below.}. Clearly, a uniform shift of all orbital positions would be such a vector. In Appendix~\ref{app:flat}, we show that, for $I$, the uniform shifts are effectively the only null vectors. For $\sigma^2_{\Omega}$, existence of additional null vectors is possible but nevertheless fine-tuned.

Let $r_{a,i}$ be a set of intra-unit-cell positions that minimize $I$. Suppose the tight-binding model is compatible with a spatial symmetry, we can apply the spatial operation to generate a new real-space realization of the tight-binding model. As shown in Appendix~\ref{app:tech}, with appropriate re-labeling, the new real-space model only differs from the original one by intra-unit-cell positions $r_{a,i}$. Further, it has the same integrated quantum metric as the original one as the spatial operations and index relabeling preserve the integrated quantum metric. Let the new intra-unit-cell positions be $r'_{a,i}$, and we see that they also correspond to a global minimum of $I$. Since the minimum of $I$ is unique up to an overall shift, we conclude that $\{r_{a,i}\} \equiv \{ r^\prime_{a,i} \}$, where $\equiv$ here denotes set equivalence up to an overall shift. This implies that the set of orbital positions that minimizes $I$ is invariant under the spatial transformation.

For $\sigma^2_{\Omega}$, if there is no additional null vector of the quadratic part of the dependence, exactly the same argument can be applied to show that it is minimized by embeddings that obey all spatial symmetries. In the fine-tuned cases where additional null vectors exist, we can generate all symmetry-related embeddings $g(\{r_{a,i}\})$, where $g \in G$ is an arbitrary element of the spatial symmetry group. Then, $\{ \tilde{r}_{a,i} \} = \frac{1}{|G|}\sum_{g \in G} g(\{ r_{a,i} \})$ is also \textit{a} global minimum (since the degeneracy arises from null \textit{vectors} of the quadratic expression) and clearly is invariant under the spatial symmetry operations. That is to say, at least one global minimum in the space of all embeddings must obey the spatial symmetries. An example of this happening is illustrated in Appendix~\ref{app:flat_B}.

Finally, let us return to the more general problem where lattice vectors $\bm{a}_i$ are allowed to vary. In Appendix~\ref{app:shape}, we show that (1) for any fixed choice of $\bm{a}_i$, $I$ is minimized by the same relative embeddings $\tilde{x}_{a}$ and $\tilde{y}_{a}$, which implies that that one can always choose lattice vectors with high symmetry when optimizing intra-unit-cell orbital positions, and (2) among all possible choices of $\bm{a}_i$, $I$ is minimized by the choice of $\bm{a}_i$ that is compatible with the symmetry of the model, so that the optimal overall embedding, when lattice vectors and intra-unit-cell positions are considered together, should obey the symmetry of the tight-binding model.

\section{Spatial symmetry and high symmetry positions}\label{sec:Wyckoff}

In the discussion in the preceding section, we have made use of only two features of spatial symmetry operations: first, a spatial symmetry operation generates a new set of orbital positions; second, the new set of orbital positions give the same quantum geometric quantities, i.e., the same $I$ and $\sigma^2_{\Omega}$. As such, the discussion applies not only to the usual point group symmetries, but also nonsymmorphic symmetries such as glide-reflection symmetries. It also applies to translation symmetries, though we have already assumed that the embedding obeys translation symmetry. A non-trivial case is the magnetic translation symmetry, which will be discussed in Sec.~\ref{sec:gauge}.

\begin{figure}
    \centering
    \includegraphics[width=0.9\linewidth]{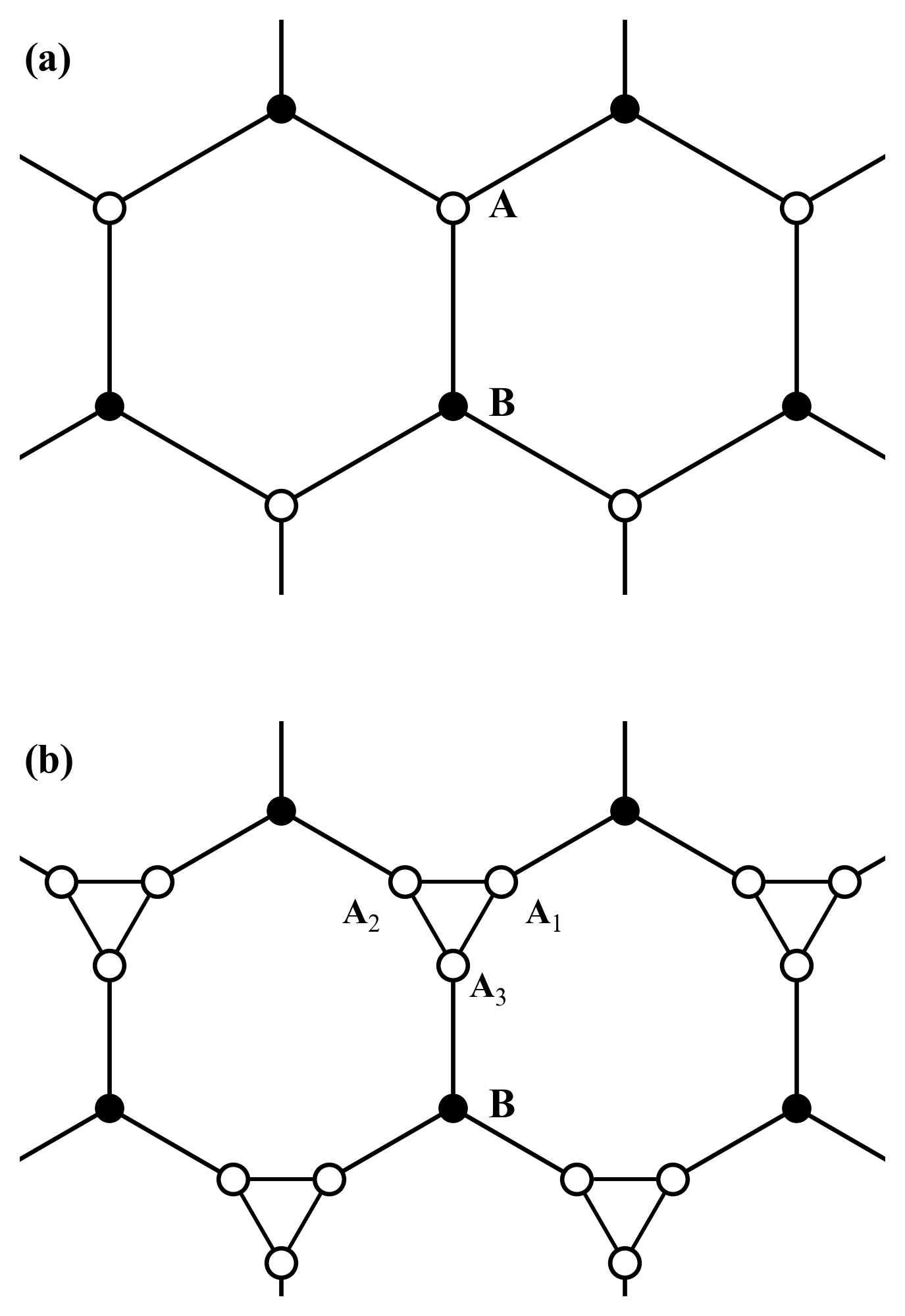}
    \caption{The honeycomb lattice (a) and a decorated honeycomb lattice (b), where each A site is replaced by a trimer. $C_3$ symmetry requires the trimer to be arranged symmetrically around the center of $C_3$-rotation, but does not uniquely specify an embedding.}
    \label{fig:honeycomb}
\end{figure}

If the embedding of a tight-binding model obeys a spatial symmetry, its orbital positions, after re-labeling, are unchanged after the spatial symmetry operation. The most obvious case is for the orbitals to be embedded at high-symmetry positions. A simple example is the honeycomb lattice, as shown shown in Fig.~\ref{fig:honeycomb}(a), where the positions of sites A and B are fixed by $C_3$ symmetry. (These positions are known as {\it special Wyckoff positions}.) 

However, it is also possible that symmetry does not fully fix orbital positions. If we replace every site of one of the two sub-lattices with three sites arranged in an appropriately placed equilateral triangle, we generate a decorated honeycomb lattice, shown in Fig.~\ref{fig:honeycomb}(b). The three sites $A_{i}$, for $i = 1,2,3$, map onto each other under $C_3$, and, as long as they are arranged in an equilateral triangle centered at the relevant high-symmetry position, the embedding is invariant under $C_3$ symmetry. As such, the side length of the equilateral triangles is not constrained by symmetry requirements. 

In the example given above, one can tune between different symmetric embeddings by continuously varying the side length of the trimer (this is indeed the only unconstrained degree of freedom up to an overall shift for fixed lattice vectors $\bm{a}_i$, if we also impose mirror symmetries). We claim that, in general, symmetric embeddings form a connected manifold. This is because, suppose $\{\bm{r}_\alpha\}$ and $\{\bm{r}'_\beta\}$ are both symmetric embeddings, i.e., for any element of the symmetry group $g \in G$, we have $\bm{r}_{\sigma(\alpha)} = g(\bm{r}_\alpha)$ and $\bm{r}'_{\sigma(\alpha)} = g(\bm{r}'_\alpha)$ while satisfying $t_{\sigma(\alpha), \sigma(\beta)} = t_{\alpha, \beta}$\footnote{An astute reader may raise the following objection: in the definition given in Section~\ref{sec:summary}, we only specified that for every $g$, there exists a corresponding permutation $\sigma$, but we do not know if it is unique, i.e. what if  $\bm{r}_{\sigma(\alpha)} = g(\bm{r}_\alpha)$ and $\bm{r}'_{\sigma'(\alpha)} = g(\bm{r}'_\alpha)$ for $\sigma \neq \sigma'$? The answer is that, for this to be possible, one must have $t_{\sigma'^{-1}(\sigma(\alpha)), \sigma'^{-1}(\sigma(\beta))} = t_{\alpha, \beta}$. This allows us to define $\bm{r}''_\alpha = \bm{r}'_{\sigma'^{-1}(\sigma(\alpha))}$. The new embedding $\{ \bm{r}''_\alpha \}$, which is merely $\{ \bm{r}'_\alpha \}$ with re-labeled indices, satisfies $\bm{r}''_{\sigma(\alpha)} = g(\bm{r}''_\alpha)$.}, then $\tilde{\bm{r}}_\alpha = \lambda \bm{r}_\alpha + (1 - \lambda) \bm{r}'_\alpha$ for $\lambda \in  \mathbb{R} $ is also a symmetric embedding, as $g$ is a linear operation up to a constant shift.

Coming back to the problem of minimizing quantum geometric properties such as $I$ and $\sigma^2_{\Omega}$, the above discussion means that, even when the symmetry does not uniquely fix the optimal embedding, it constrains the possible choices to a connected sub-manifold of all embeddings.

\section{Gauge transformation and time-reversal symmetry}\label{sec:gauge}

In the preceding discussions, we have made no allowance for gauge transformation. Many models, however, only manifest certain spatial symmetries if we allow for a gauge transformation, as defined in Eq.~\ref{eq:gauge}. For example, consider the Qi-Wu-Zhang model~\cite{qi_topological_2006}, given by
\begin{equation}
H(\bm{k})=\sin k_x\,\sigma_x+\sin k_y\,\sigma_y+\left(m+\cos k_x+\cos k_y\right)\sigma_z,
\end{equation}
for $k_x \in [0, 2\pi)$, $k_y \in [0, 2\pi)$ and some constant $m$. The $C_4$ symmetry that acts as
\begin{equation}
    C_4: \ (k_x, k_y) \rightarrow (-k_y, k_x).
\end{equation}
This is evidently \textit{not} a symmetry of the Hamiltonian, as
\begin{equation}
    H(-k_y, k_x) = -\sin k_y\,\sigma_x+\sin k_x\,\sigma_y+\left(m+\cos k_x+\cos k_y\right)\sigma_z,
\end{equation}
which is in general not equal to $H(k_x, k_y)$. However, it is easy to show that the $C_4$ spatial rotation, when combined with the gauge transformation given by
\begin{equation}
U = e^{-i\frac{\pi}{4}\sigma_z},    
\end{equation}
is a symmetry of the Hamiltonian.

    We have so far shown that the embedding that minimizes quantum geometric quantities $I$ or $\sigma^2_{\Omega}$ should obey all compatible spatial symmetries. The question that arises now is whether the statement applies when the spatial symmetry comes with a gauge transformation. In demonstrating the statement without gauge transformation, we have used the fact that a spatial transformation does not alter the quantum geometric quantities. To answer the new question in the affirmative, we need to show that gauge transformation also does not alter the mentioned quantum geometric properties\footnote{When people usually say ``the quantum metric tensor is gauge-invariant'', they are referring to gauge transformations of the form $\ket{u(\bm{k})} \rightarrow e^{i\phi(\bm{k})}\ket{u(\bm{k})}$, which is \textit{not} what we mean by ``gauge transformation'' here.}~\footnote{One may think that the claim is trivial, in that, since the spatial operation combined with gauge transformation is a symmetry of the Hamiltonian, and the spatial operation itself does not change the quantum geometric properties, the gauge transformation surely does not alter the quantum geometric properties and there is nothing left to show. However, this is missing the point of the argument. We want to show, for an arbitrary embedding which may not obey the combined symmetry, a spatial transformation when combined with gauge transformation does not modify the quantum geometric properties. To do so, we need to show that the gauge transformation does not alter quantum geometric properties.}. In the given example, the gauge transformation has a particularly simple structure, namely
\begin{equation}
    \hat{c}_{\bm{R},a} \rightarrow \hat{c}_{\bm{R},a}e^{i\phi_{a}}.
\end{equation}
That is, the additional phase is independent of the unit cell $\bm{R}$. Such a transformation is equivalent to 
\begin{equation}
    \tilde{u}_a(\bm{k}) =  u_a(\bm{k})e^{i\phi_{a}},
\end{equation}
and it is easy to see that such a transformation keeps the quantum geometric tensor $Q_{ij}$ invariant, which naturally implies that both $I$ and $\sigma^2_{\Omega}$ are invariant.

However, not all physically interesting gauge transformations can be written down in such a simple form. For example, consider the Hofstadter model with $1/q$-flux per plaquette, given by
\begin{equation}
H =
-t \sum_{m,n}
\left(
c^{\dagger}_{m+1,n} c_{m,n}
+
e^{i \frac{2\pi}{q} m}
c^{\dagger}_{m,n+1} c_{m,n}
+ \mathrm{h.c.}
\right).
\end{equation}
The magnetic translation operators are defined by

\begin{align}
T_x \, \hat{c}_{m,n} \, T_x^{-1}
=&
e^{-i \frac{2\pi}{q} n}
\hat{c}_{m+1,n}\\
T_y \, \hat{c}_{m,n} \, T_y^{-1}
=&
\hat{c}_{m,n+1}.
\end{align}
In this gauge, $T_y$ is just an ordinary translation, while $T_x$ is a translation combined with a gauge transformation 
\begin{equation}
    U\hat{c}_{m,n}U^{-1} = e^{-i\frac{2\pi}{q}n}\hat{c}_{m,n}.
\end{equation}
The {\it magnetic} unit cell of the model under the given gauge consists of $q \times 1$ plaquettes, and translation by a magnetic unit cell is a symmetry without any gauge transform.  However, translation by the elementary unit cell requires gauge transform to be a symmetry.   Indeed, the gauge transformation written above involves a phase-factor that is unit-cell \textit{dependent}. As such, we need a new argument to show that this kind of gauge transformations also do not alter quantum geometric properties.

One may note that the phase dependence on unit-cell position for the above-written gauge transformation is linear. To generalize, we may consider the following type of gauge transformation,
\begin{equation}\label{eq:linear_gauge}
    \hat{c}_{\bm{R},a} \rightarrow \hat{c}_{\bm{R},a}e^{i(\bm{k}_0\cdot \bm{R} + \phi_a)},
\end{equation}
for some constant vector $\bm{k}_0$. Under such a gauge transformation, the wavefunction transforms as
\begin{equation}
    \tilde{u}_a(\bm{k} + \bm{k}_0) = e^{i(-\bm{k}_0 \cdot \bm{r}_a + \phi_a)}u_a(\bm{k}).
\end{equation}
Clearly, under such a transformation,
\begin{align}
    \tilde{g}_{ij}(\bm{k}) &= g_{ij}(\bm{k} - \bm{k}_0) \\
    \tilde{\Omega}(\bm{k}) &= \Omega(\bm{k} - \bm{k}_0),
\end{align}
and the integrated quantities such as $I$ and $\sigma^2_{\Omega}$ are not affected.

Returning to the Hofstadter model with $1/q$ flux per plaquette, with a given gauge, a valid embedding only needs to preserve the periodicity of magnetic unit cells, with arbitrary positions for the $q$ sites within the magnetic unit cell. What we have shown here is that, for the model to have minimal $I$ or $\sigma^2_{\Omega}$, the embeddings of orbitals within the magnetic unit cell must be such that they are consistent with the magnetic translation symmetry, i.e., placed on a regular grid. Such an arrangement is usually the default choice when studying such models, but it is noteworthy that this choice is also the optimal one for quantum geometric properties.

Actually, in practice, sometimes one encounters gauge transformations that do not follow the form set out in Eq.~\ref{eq:linear_gauge}. For example, the Hofstadter model has a $C_4$ symmetry, but the accompanying gauge transformation generally depends on the unit cell position in a non-trivial manner. This is because, for example, suppose the original magnetic unit cell is $4 \times 1$, a pure $C_4$ rotation without gauge transformation maps the magnetic unit cell to $1\times 4$. For the combined operation to be a symmetry of the Hamiltonian, the gauge transformation must modify the periodicity of the model. In Appendix~\ref{app:gauge}, we demonstrate that arbitrary gauge transformations, assuming that they do not completely remove the periodic properties of the Hamiltonian, do not alter the $I$. For $\sigma^2_{\Omega}$, the same is true provided some limit on increasing the size of the magnetic unit cell is obeyed, which does not pose a problem for our purposes.\footnote{This is because we are interested in gauge transformations that follow $t_{\sigma(\alpha), \sigma(\beta)} = t_{\alpha, \beta}e^{-i(\phi_\alpha - \phi_\beta)}$, with $\sigma$ being the permutation of orbitals associated with some spatial symmetry. Therefore, gauge transformations relevant to our purposes should conserve the area of the magnetic unit cell.} Furthermore, it is also easy to see that the (spinless) time-reversal operation has no effect on $I$ or $\sigma^2_{\Omega}$. Therefore, following the same argument as in Sec.~\ref{sec:proof}, we see that, in order to minimize $I$ or $\sigma^2_{\Omega}$, the real-space realization must obey all spatial symmetries \textit{up to gauge transformation and possible time-reversal operation}.

\section{Corollaries and extensions}\label{sec:corollaries}

(1) In the context of fractional Chern insulators, one commonly invoked condition is the ideal band condition, which can be written as~\cite{roy_band_2014}
\begin{equation}\label{eq:trace}
    I \equiv \int_{\text{BZ}}d^2\bm{k} \tr[g] = 2\pi |C|,
\end{equation}
where $C$ is the Chern number. Since $2\pi |C|$ is also the lower bound of $I$ for any Bloch band with Chern number $C$~\cite{roy_band_2014}, satisfying Eq.~\ref{eq:trace} necessarily implies that the embedding has been chosen to minimize $I$. This immediately implies that the embedding of an ideal band must obey all symmetries of the underlying tight-binding model. For example, the Kapit-Mueller model~\cite{kapit_exact_2010} is similar to the Hofstadter model but with the hopping amplitudes carefully chosen such that the wavefunctions form a lattice version of the Landau levels, which immediately implies that it is an ideal band~\cite{wang_exact_2021}. This model has the same symmetry as the Hofstadter problem, including magnetic translation symmetries and $C_4$ rotation symmetry up to gauge, and we now see that it is not a coincidence that its ``idealness'' is achieved when the embedding is chosen to obey these symmetries. 

At this point, we would like to remind the reader that in the literature sometimes one uses a Fourier convention such that the Hamiltonian $H(\bm{k})$ is periodic in momentum space (called convention I in Ref.~\cite{fruchart_parallel_2014}), which is equivalent to placing all orbitals of one unit cell at the same location in our framework. If one adopts this convention and calculates the quantum geometric properties without corrections, most ideal bands will not appear ideal. This is another demonstration that such a convention is inappropriate for the purpose of calculating quantum geometric measures. 

Finally, as we note in Appendix~\ref{app:no_symmetry}, it is possible to construct ideal bands without any spatial symmetry (other than the necessary translation symmetries for the Bloch theorem to hold), so the symmetry argument does not always constrain the embedding of an ideal band. 

(2) In proving that gauge transformation does not change quantum geometric properties like $I$ or $\sigma^2_{\Omega}$ (Appendix~\ref{app:gauge}), we see that different choices of magnetic unit cells do not affect these quantum geometric properties. For example, for a Hofstadter problem with $1/4$-flux per plaquette, provided the orbitals are arranged on the same square grid, a $2 \times 2$ magnetic unit cell and a $4 \times 1$ magnetic unit cell yield the same quantum geometric properties.

(3) So far we have focused on the case of a single isolated band. For a group of bands, the quantum geometric tensor is defined as
\begin{equation}\label{eq:Q_multibands}
    Q_{ij} = \sum_n \braket{\partial_{k_i}u_n(\bm{k})|Q(\bm{k})|\partial_{k_j}u_n(\bm{k})},
\end{equation}
where $Q(\bm{k}) = 1 - \sum_n\ket{u_n(\bm{k})}\bra{u_n(\bm{k})}$. We can again define the quantum metric tensor and the (abelian) Berry curvature by the real and imaginary parts of the quantum geometric tensor. It is straightforward to show that the statements about minimizing $I$ and $\sigma^2_{\Omega}$ are also applicable here.

(4) The two quantities we have focused on in this work, namely the variance of Berry curvature $\sigma^2_\Omega$ and the integrated quantum metric $I$, share the key property of depending on (intra-unit-cell) orbital positions as a polynomial up to the second order. The argument we have presented can largely be applied to other quantum geometric measures provided they also depend on orbital positions only up to the second order (additional care may be required to determine the uniqueness of the minima and dependence on lattice vectors). For example, since the Berry curvature depends on the orbital position only up to linear order, our argument applies to other quantities that can be constructed from a quadratic expression of Berry curvature (provided the quantities are lower-bounded). As an illustration, in Appendix~\ref{app:wannier}, we consider the spread of the maximally localized Wannier functions for an isolated band, and we show that it is minimized (on top of minimization over Bloch phases) by symmetric choice of orbital positions. 

(5) As one can think of a continuum model as the limiting case of a tight-binding model with densely placed orbitals, much of the discussion in this work can be extended to continuum models, provided that changing orbital positions is understood as a diffeomorphism on the real space, i.e., for some (bijective and unit-cell periodic) map $\bm{r} \rightarrow \bm{r}'$, we transform the Bloch wavefunctions as
\begin{equation}
    \tilde{u}_{\bm{k}}(\bm{r}') = e^{-i\bm{k} \cdot (\bm{r}' - \bm{r})}\sqrt{\left|\frac{\partial \bm{r}}{\partial \bm{r}'}\right|}u_{\bm{k}}(\bm{r}).
\end{equation}
For continuum models, symmetry constrains are usually not sufficient to entirely fix the optimal embedding (with the notable exception of Landau levels, where continuous translation symmetries are strong enough to fully constrain the embedding), so that even if one has chosen a symmetric embedding by default, the quantum geometric properties may still be open to further optimization by diffeomorphism.

(6) Besides the integrated quantum metric, another quantum geometric measure that is sometimes discussed in the literature is the ``quantum volume'', defined as
\begin{equation}
    \mathcal{V} = \int_{\text{BZ}} d^2\bm{k} \ \sqrt{\det[g]}.
\end{equation}
While its physical application is less common than the integrated quantum metric discussed so far, it has the nice mathematical interpretation of being the volume of the Bloch wavefunctions in the Hilbert space under the Fubini-Study metric. Unlike the integrated quantum metric, the quantum volume is not a low-degree polynomial of the orbital positions. While the same symmetry arguments can be used to show that symmetric orbital positions correspond to stationary points of the quantum volume, in Appendix~\ref{app:quantum_volume}, we show that the global minimum of the quantum volume does not always correspond to a symmetric embedding.

\section{Conclusion}\label{sec:conclusion}

In this work, we have studied the problem of minimizing the variance of Berry curvature and the integrated quantum metric over real-space orbital positions (embeddings) for a tight-binding model. Our central conclusion is that the optimal choice of orbital positions should obey all the spatial symmetries that the tight-binding model is compatible with, up to gauge transformation and possible time-reversal. For some cases, this symmetry requirement uniquely fixes the optimal positions of the orbitals, and places them at special Wyckoff positions. For other cases, the symmetry requirement does not fully pin down the orbital positions. Instead the symmetry-allowed embeddings form a connected sub-manifold of all possible embeddings. We also investigate the degeneracy of minima for the variance of Berry curvature and integrated quantum metric respectively. We find that the minimum of the integrated quantum metric is unique up to an overall shift, but we establish non-trivial cases where the minima are degenerate for the variance of Berry curvature. Corollaries of our results include that the conventional embedding of the Hofstadter model actually minimizes the variance of Berry curvature and the integrated quantum metric over all possible rearrangement of orbitals within the magnetic unit cell, and that the embedding of an ideal band must obey all the symmetries that the underlying tight-binding model is compatible with.

We have so far treated the formulation of a tight-binding model as agnostic to the orbital positions. This approach is justified, as many physical problems one may be interested in, such as the stability of fractional Chern insulators, can be fully formulated without reference to orbital positions (termed ``geometry-independent'' in Ref.~\cite{simon_contrasting_2020}). Nevertheless, if a tight-binding model is derived from a particular physical system, there would be a set of physical orbital positions. As the hopping amplitudes are related to the physical distance, the physical orbital positions would normally have the same symmetry as the tight-binding model. If symmetry fully constrains the orbital positions, the physical orbital positions would then be the only set of symmetric orbital positions (up to an overall shift as usual) and therefore necessarily minimize the variance of Berry curvature and the integrated quantum metric. However, in the cases where symmetry does not uniquely pin down the embedding, it is possible that the physical orbital positions differ from the embedding that minimizes the these properties. For geometry-independent observables, the physical choice would then introduce information that is not contained in the tight-binding model itself, making it less appropriate.

{\bf Acknowledgements:} The authors acknowledge helpful conversations with Jonah Herzog-Arbeitman.   This work was partially supported by a Leverhulme Trust International Professorship (Grant Number
LIP-202-014, Z.W.) and the UK Engineering and Physical Sciences Research Council EPSRC (EP/X030881/1, S.H.S.).

During the final stage of our work, we became
aware of Refs.~\cite{li_quantum-geometric_2026, to_appear}, whose results overlap with those presented in Appendix~\ref{app:no_symmetry}.

\appendix

\section{Generation of new orbital positions}\label{app:tech}

We would like to consider how spatial operations transform the embedding of a tight-binding model, where the spatial operation is not necessarily a symmetry of the real-space model. On the most basic level, suppose we have some embedding $\bm{r}_\alpha$, a spatial operation $g$ generates a new embedding $\tilde{\bm{r}}_\alpha \equiv g(\bm{r}_\alpha)$. This is in general not very interesting, as the new embedding alters both the lattice vectors and the intra-unit-cell positions. However, there are circumstances where the spatial operation generates an embedding which, after appropriate re-labeling, only differs from the original one by intra-unit-cell positions. To start, we consider a tight-binding model that is compatible with a spatial symmetry $g$ such that a symmetric realization is given by
\begin{equation}
    \bm{r}^{\text{(sym)}}_{m,n,a} = m\bm{a}_1 + n\bm{a}_2 + \bm{r}_a,
\end{equation}
which obeys
\begin{equation}
    \bm{r}^{\text{(sym)}}_{\sigma(m,n,a)} = g(\bm{r}^{\text{(sym)}}_{m,n,a}),
\end{equation}
where $\sigma$ is the permutation discussed in Section~\ref{sec:summary}, i.e., it satisfies $t_{\sigma(\alpha),\sigma(\beta)} = t_{\alpha, \beta}$ (in this Appendix, we are not yet introducing gauge transformations).

Now, let us consider a different embedding with the same lattice vectors but different intra-unit-cell positions i.e., 
\begin{equation}
    \bm{r}_{m,n,a} = m\bm{a}_1 + n\bm{a}_2 + \bm{r}_a + \delta\bm{r}_a.
\end{equation}
Applying the spatial operation, we obtain
\begin{equation}
    g(\bm{r}_{m,n,a}) = g(\bm{r}^{\text{(sym)}}_{m,n,a}) + g(\delta\bm{r}_a).
\end{equation}
Instead of naively defining the new positions of orbital $(m,n,a)$ to $g(\bm{r}_{m,n,a})$, we re-label the indices such that
\begin{equation}
    \tilde{\bm{r}}_{\sigma(m,n,a)} = g(\bm{r}_{m,n,a}).
\end{equation}
This re-labeling does not physically alter the real-space model, since the hopping amplitudes have been assumed to be invariant under the permutation. Since a permutation is invertible, we can also write
\begin{equation}
    \tilde{\bm{r}}_{m,n,a} = g(\bm{r}_{\sigma^{-1}(m,n,a)}).
\end{equation}
Let $(m',n',a') = \sigma^{-1}(m,n,a)$. In general, $m'$ depends not only on $m$, but also on $n$ and $a$. However, due to translation symmetry, $a'$ depends only on $a$, so that, with a slight abuse of notation, we may write $a' = \sigma^{-1}(a)$. This implies
\begin{equation}
    \tilde{\bm{r}}_{m,n,a} = g(\bm{r}^{\text{(sym)}}_{\sigma^{-1}(m,n,a)} + \delta\bm{r}_{\sigma^{-1}(a)}).
\end{equation}
By definition, one has
\begin{equation}
    g(\bm{r}^{\text{(sym)}}_{\sigma^{-1}(m,n,a)}) = \bm{r}^{\text{(sym)}}_{m,n,a},
\end{equation}
so that
\begin{equation}
    \tilde{\bm{r}}_{m,n,a} = \bm{r}_{m,n,a} -\delta \bm{r}_{a} + g(\delta\bm{r}_{\sigma^{-1}(a)}).
\end{equation}
That is, the new embedding differs from the original one only by intra-unit-cell positions.

\section{Uniqueness of minima}

\subsection{Integrated quantum metric}\label{app:flat}

In this section, we show that, excluding trivially removable orbitals, the only (two) flat directions (i.e. combinations of $\{ r_{a,i} \}$ that are not all zero and keep the quantum geometric quantity unchanged) for the integrated quantum metric are the uniform shifts in the orbital positions.

The diagonal components of the quantum metric tensor are given by
\begin{equation}
    g_{ii} = \sum_a \partial_{k_i}u^*_a(\bm{k}) \partial_{k_i}u_a(\bm{k}) - \left|\sum_a u^*_a(\bm{k})\partial_{k_i}u_a(\bm{k})\right|^2 
\end{equation}
After the change in embedding, Bloch wavefunctions transform as
\begin{equation}\label{eqn:transform}
    u_a(\bm{k}) \rightarrow e^{-i \bm{k} \cdot \bm{r}_a}u_a(\bm{k}).
\end{equation}
By direct substitution, one can show that $g_{ii}$ transforms as
\begin{equation}\label{eq:gii}
    \begin{split}
        \tilde{g}_{ii} \ = \ &g_{ii}  \sum_a\left[2i|u_a|^2 \braket{\partial_{k_i}u|u} - i\partial_{k_i} u^*_a u_a + i u^*_a \partial_{k_i}u_a\right]r_{a,i} \\ +& \sum_a(r_{a,i})^2|u_a|^2 - \left(\sum_a r_{a,i}|u_a|^2\right)^2.
    \end{split}
\end{equation}     
We are interested in minimizing the quantity
\begin{equation}
    I(\{ \bm{r}_a \}) \equiv \int d^2\bm{k} \tr[\tilde{g}] \equiv \int d^2\bm{k} (\tilde{g}_{xx} + \tilde{g}_{yy}).
\end{equation}
As discussed in the main text, $I(\{ \bm{r}_a \})$ is a polynomial of $r_{a,i}$ up to quadratic order, and we want to know whether there is any flat direction in the $r_{a,i}$-space. We claim that, assuming there is no trivially removable orbitals (a trivially removable orbital is one where $u_a(\bm{k}) = 0$ for all $\bm{k}$), the only flat directions are the uniform shifts of all orbitals.

First, since $\tilde{g}_{ii}$ (no sum) only depends on $r_{a,i}$, we can discuss the flat directions of $\int d^2\bm{k} \ \tilde{g}_{xx}$ in the space of $x_a$ (and the equivalent situation for $\int d^2\bm{k} \ \tilde{g}_{yy}$) separately. Also, for a quadratic expression to be constant, both the coefficients of the linear and the quadratic part must vanish. As such, we can just focus on the flat directions of the quadratic part of the expression, which is
\begin{equation}
\begin{split}
       &X(\{ x_a\})  \equiv \\ &\int d^2 \bm{k} \left[\sum_a(x_a)^2|u_a(\bm{k})|^2 - \left(\sum_a x_a|u_a(\bm{k})|^2\right)^2\right],
\end{split}
\end{equation}
and similarly for $\int d^2\bm{k} \ \tilde{g}_{yy}$. A flat direction of the quadratic form is one where $X(x_a) = 0$ for $x_a$ not all zero. One can show that the integrand is nowhere negative, as the Cauchy–Schwarz inequality implies that
\begin{equation}
\left(\sum_a |u_a|^2x_a\right)^2
\le
\left(\sum_a |u_a|^2\right)
\left(\sum_a |u_a|^2 x_a^2\right),
\end{equation}
and $\sum_a |u_a|^2 = 1$.Therefore, for $X(x_a) = 0$ to hold, the integrand must be zero identically for each $\bm{k}$, i.e., the Cauchy–Schwarz inequality must be saturated at each $\bm{k}$. It is easy to prove that the inequality is satisfied if and only if \textit{all $x_a$ with non-zero $u_a$ are equal.} 

Of course, all $x_a$ being equal saturates the inequality. This simply corresponds to a uniform shift in the embedding. Also, if $u_a(\bm{k}) = 0$ for all $\bm{k}$, we can choose $\bm{r}_a$ to be anything and it does not affect the quantum geometry. This corresponds to a trivially removable orbital, which we will exclude from now on. Are there any other possibilities? The answer is no, assuming the band is not singular. 

To demonstrate this more rigorously, we define an equivalence relation on the set of orbitals such that two orbitals $a$ and $b$ are ``equivalent'' if there is some $\bm{k}$ such that $u_a(\bm{k})$ and $u_{b}(\bm{k})$ are both non-zero\footnote{This is a sufficient but not necessary condition. It may be that $a$ and $b$ are equivalent only through an intermediary $c$ such that both $u_a$ and $u_c$ are non-zero for some $\bm{k}$, and $u_b$ and $u_c$ are non-zero for some other $\bm{k}$.}. Based on the preceding argument, for $X(\{ x_a\}) = 0$ to hold, two orbitals $a,b$ in the same equivalence class must have $x_a = x_b$. On the other hand, each equivalence class corresponds to some $\bm{k}$-space region where their weights do not vanish. If there is more than one equivalence class, there must be a situation where near their boundary in $\bm{k}$-space, the wavefunctions have non-zero weights only on one set of orbitals on one side of the boundary, and have non-zero weights only on another mutually exclusive set of orbitals on the other side of the boundary. The wavefunction must therefore be discontinuous at the boundary, which we do not allow. This shows that there can only be one equivalence class of orbitals, and, as we have shown, $X(\{ x_a\}) = 0$ would then require $x_a = x_b$, i.e., a uniform shift. The same discussion, of course, can be applied to $\int d^2\bm{k} \ \tilde{g}_{yy}$.

At this point, we should note that the discussion above is meant to be applied to the case of a single band, as assumed throughout most of this work. Suppose we are considering the quantum metric of two (or more) bands together, then, if the two bands are supported on mutually exclusive sets of orbitals (say, they have different flavor indices), the relative positions between these two sets of orbitals clearly do not affect the total integrated quantum metric. 

\subsection{Variance of Berry curvature}\label{app:flat_B}

Now we turn our attention to the variance of Berry curvature. Under a change of embedding, the Berry curvature transforms as
\begin{equation}
    \Omega(\bm{k}) \rightarrow \Omega(\bm{k}) - \sum_a \epsilon^{ij}r_{a,i}\partial_{k_j}|u_a(\bm{k})|^2.
\end{equation}
For a certain combination of $r_{a,i}$ to be a flat direction, the necessary and sufficient condition is
\begin{equation}
    \sum_a \epsilon^{ij}r_{a,i}\partial_{k_j}|u_a(\bm{k})|^2 = 0 \quad \text{for all} \ \bm{k}.
\end{equation}
Clearly, a uniform shift is a flat direction as $\sum_a \partial_{k_i}|u_a(\bm{k})|^2= 0$ for $i = x,y$, but this is not the only possibility. For example, let $S$ denote a proper subset of the intra-unit-cell atomic basis, and $\sum_{a \in S}|u_a(\bm{k})|^2$ is a constant in $\bm{k}$, then a uniform shift of all orbitals within $S$ keeps $\Omega(\bm{k})$ constant for all $\bm{k}$ and therefore preserves the variance of Berry curvature.

To illustrate this more clearly, we construct the following model. Consider some generic tight-binding Bloch Hamiltonian given by $\hat{H}_0 = \sum_{\alpha,\beta} t_{\alpha, \beta} \hat{c}^\dagger_\alpha \hat{c}_\beta$, and assume its lowest energy band is isolated. Now, consider the following ``doubled'' version of the Hamiltonian with some ``external field'' $h > 0$, such that
\begin{equation}
    \hat{H} = \sum_{\alpha, \beta} t_{\alpha, \beta} (\hat{c}^\dagger_{\alpha,1} \hat{c}_{\beta,1} + \hat{c}^\dagger_{\alpha,2} \hat{c}_{\beta,2}) - h\sum_\alpha (\hat{c}^\dagger_{\alpha,1}\hat{c}_{\alpha,2} + \hat{c}^\dagger_{\alpha,2}\hat{c}_{\alpha,1}).
\end{equation}
If the Bloch wavefunction of the lowest band of $\hat{H}_0$ is $\ket{\psi^0(\bm{k})}$, then that of the doubled Hamiltonian $\hat{H}$ is
\begin{equation}
    \ket{\psi(\bm{k})} =  \ket{\psi^0(\bm{k})} \otimes \frac{1}{\sqrt{2}}
\begin{pmatrix}
1 \\
1
\end{pmatrix}.
\end{equation}
Assuming the corresponding orbitals of the two copies are placed at the same spatial location, i.e., $\bm{r}_{\alpha,2} = \bm{r}_{\alpha,1}$, we have
\begin{equation}
    \ket{u(\bm{k})} =  \ket{u^0(\bm{k})} \otimes \frac{1}{\sqrt{2}}
\begin{pmatrix}
1 \\
1
\end{pmatrix}.
\end{equation}
If we choose $\bm{r}_{\alpha,2} = \bm{r}_{\alpha,1} + \delta \bm{r}$ for some constant $\delta\bm{r}$ instead, we have 
\begin{equation}
    \ket{\tilde{u}(\bm{k})} =  \ket{u^0(\bm{k})} \otimes \frac{1}{\sqrt{2}}
\begin{pmatrix}
1 \\
e^{-i\bm{k}\cdot \delta\bm{r}}
\end{pmatrix}.
\end{equation}
It is straightforward to demonstrate that $\ket{u(\bm{k})}$ and $\ket{\tilde{u}(\bm{k})}$ produce the same Berry curvature at every $\bm{k}$ so that their variances of Berry curvature are the same. Nevertheless, the difference of the two embeddings is not a constant shift of \emph{all} orbitals.

\begin{figure}[h]
    \centering
    \includegraphics[width=\linewidth]{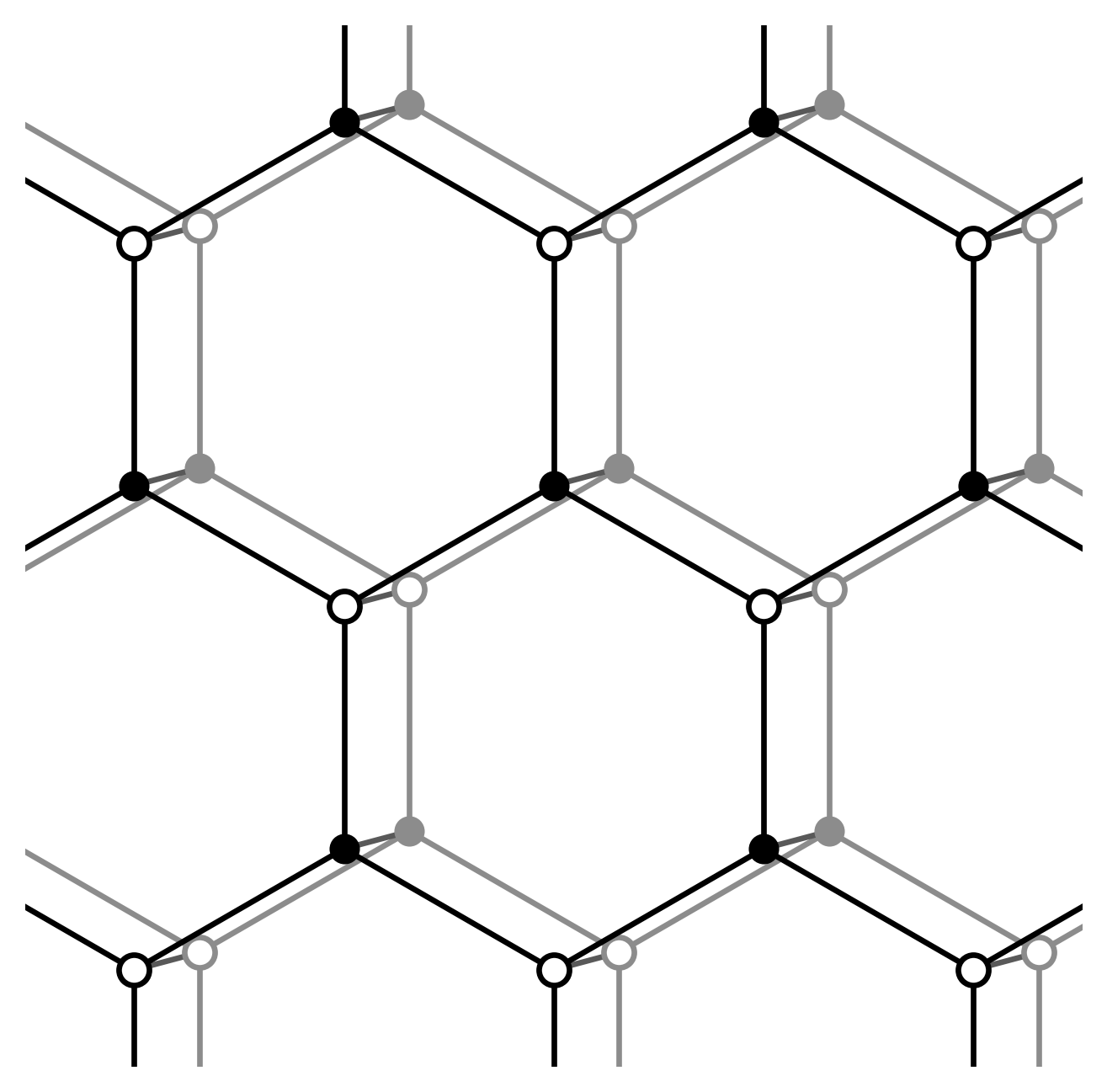}
    \caption{Two honeycomb lattices with an offset and bonds in-between. $C_3$-symmetry for the entire lattice is only restored if the offset is zero. }
    \label{fig:double_lattice}
\end{figure}

In order to relate the above discussion to the issue of symmetry, let us consider $\hat{H}^0$ to be the Haldane model, which has $C_3$-symmetry. We use the procedure outlined above to construct a ``doubled'' version of the Haldane model~\cite{haldane_model_1988}, as illustrated in Fig.~\ref{fig:double_lattice} (for simplicity, the next-nearest neighboring hoppings are not drawn explicitly).  The real-space model loses the $C_3$-symmetry if there is a finite spatial offset between the two honeycomb lattices. As we have argued at the end of Section~\ref{sec:proof}, there exists a global minimum of the variance of Berry curvature with symmetric embedding. Since there is only one symmetric embedding of this model, corresponding to no offset between the two copies of honeycomb lattices, this embedding must be a global minimum. However, as we have argued, the offset does not affect the variance of Berry curvature. This means that, though not obeying $C_3$ symmetry, embedding with a finite offset is also a global minimum of the variance of Berry curvature.

Finally, we note that, even though the variance of Berry curvature does not depend on the offset, the integrated quantum metric does depend on the offset, so the global minimum of integrated quantum metric remains unique up to overall translation.

\section{Shape of the unit cell}\label{app:shape}

In this section, we consider the implication of changing the shape of the unit cell for the quantum metric tensor. Let $\bm{a}_i$ for $i = 1,2$ denote a certain choice of lattice vectors. The position of each orbital can then be written as $\bm{r}_a = \tilde{x}_{a}\bm{a}_1 + \tilde{y}_a\bm{a}_2$ for a pair of real numbers $(\tilde{x}_{a}, \tilde{y}_{a})$. Suppose two real-space realizations differ by $\bm{a}_i$ but have the same $(\tilde{x}_{a}, \tilde{y}_{a})$, we say that the two configurations have the same \textit{relative} embedding. 

Let us define BZ-integrated components of the quantum metric tensor as
\begin{equation}
    \bar{g}_{ij} = \int_{\text{BZ}} d^2\bm{k} \ g_{ij}(\bm{k}).
\end{equation}
A deformation of the lattice vectors can be denoted by a $2 \times 2$ unimodular\footnote{Here we assume, without loss of generality, that the lattice vector change does not affect the area of unit cell.} matrix $M$ such that $\bm{r} \rightarrow M\bm{r}$. The quantum metric, given the same \textit{relative} embedding, transforms as
\begin{equation}
    \bar{g} \rightarrow M\bar{g}M^T.
\end{equation}

Referring to Eq.~\ref{eq:gii}, one observes that $\bar{g}_{xx}$ only depends on the $x$-coordinates of the embedding, similarly for $\bar{g}_{yy}$. This means that one can simultaneously minimize both $\bar{g}_{xx}$ and $\bar{g}_{yy}$, which of course then minimizes their sum, i.e., $\text{tr}[\bar{g}]$. We have seen that, for each choice of $\bm{a}_{i}$, there is a particular set of embeddings (always up to uniform translation) that minimizes $\text{tr}[\bar{g}]$. We claim that, for any choice of $\bm{a}_{i}$, the relative embedding that minimizes $\text{tr}[\bar{g}]$ is the same, as proved in the following paragraphs.

Any two choices of lattice vectors are related by some transformation $M$, which can be decomposed as a re-scaling sandwiched between rotations, i.e.,
\begin{equation}
M
=
R_\theta
\begin{pmatrix}
\alpha & 0\\
0 & \alpha^{-1}
\end{pmatrix}
R_\phi,
\qquad \alpha>0.
\end{equation}

\begin{equation}
R_\theta=
\begin{pmatrix}
\cos\theta & -\sin\theta\\
\sin\theta & \cos\theta
\end{pmatrix},
\qquad
R_\phi=
\begin{pmatrix}
\cos\phi & -\sin\phi\\
\sin\phi & \cos\phi
\end{pmatrix}.
\end{equation}
From right to left, this encodes a $\phi$-rotation, an $\alpha$-rescaling and another $\theta$-rotation. Clearly, \textit{a rotation does not change the optimal relative embedding.} Therefore, we only need to show that re-scaling also does not affect the optimal relative embedding in order to demonstrate that optimal relative embedding is the same for all choices of basis vectors. For a relative embedding $\tilde{x}_a, \tilde{y}_a$, the quantum metric tensors before and after the rescaling are related by:

\begin{align}
     \bar{g}^\alpha_{xx}(\tilde{x}_a, \tilde{y}_a) &= \alpha^2 \bar{g}_{xx}(\tilde{x}_a, \tilde{y}_a), \\
     \bar{g}^\alpha_{yy}(\tilde{x}_a, \tilde{y}_a) &=  \frac{1}{\alpha^2} \bar{g}_{yy}(\tilde{x}_a, \tilde{y}_a).
\end{align}

As the lattice vectors are not guaranteed to be parallel to $x$ or $y-$directions, $\bar{g}_{xx}$ can depend on both $\tilde{x}_a$ and $\tilde{y}_a$. Nevertheless, as argued above, there is still some embedding that \textit{simultaneously} minimizes both $\bar{g}_{xx}$ and $\bar{g}_{yy}$, and we call this (relative) embedding  $(\tilde{x}_{a,\text{min}}, \tilde{y}_{a,\text{min}})$. Clearly, such an embedding also simultaneously minimizes
\begin{align}
  \text{tr}[\bar{g}(\tilde{x}_a, \tilde{y}_a)] &= \bar{g}_{xx}(\tilde{x}_a, \tilde{y}_a) + \bar{g}_{yy}(\tilde{x}_a, \tilde{y}_a) \\
\text{tr}[\bar{g}^\alpha(\tilde{x}_a, \tilde{y}_a)] &= \alpha^2 \bar{g}_{xx}(\tilde{x}_a, \tilde{y}_a) +  \frac{1}{\alpha^2} \bar{g}_{yy}(\tilde{x}_a, \tilde{y}_a),
\end{align}
so that it is the optimal relative embedding for both the original and re-scaled lattice vectors. 

In general, suppose a hopping model is compatible with some spatial symmetry, it may be necessary that, besides choosing the relative embedding appropriately, one also needs to choose appropriate lattice vectors for the symmetry to be realized. For example, for a model to realize a $C_3$ symmetry, the lattice must be hexagonal. From what we have shown, the optimal relative embedding is the same regardless of the choice of lattice vectors. Therefore, in order to determine it based on symmetry arguments, we can always choose the lattice vectors to be compatible with the symmetry so that the arguments in Sec.~\ref{sec:proof} is applicable, which in turn determines the optimal embedding for arbitrary choice of lattice vectors by keeping the same relative embedding.

We can go one step further. In the following, we would like to compare different choices of lattice vectors and show that \textit{lattice vectors that are compatible with the symmetry are optimal}. We will always assume that the relative embedding is the common optimal relative embedding, drop the dependence $(\tilde{x}_a, \tilde{y}_a)$, and write $\bar{g}$ simply to refer to the quantum metric tensor with the optimal embedding for a given choice of lattice vectors. Since the issue of choosing symmetry-compatible lattice vectors only arises for $C_3$ (including $C_6$), $C_4$, ordinary and glide reflection symmetries, we will simply discuss them case by case.

Consider hopping models that are consistent with $C_4$ or $C_3$ symmetry, and suppose we have chosen the corresponding Bravais lattice (square or hexagonal). Then, the trace of the quantum metric tensor is minimized when the embedding also preserves the $C_4$ or $C_3$ symmetry. We call the quantum metric tensor $\bar{g}^0$, and due to the $C_4$ or $C_3$ symmetry, it must be proportional to the identity. The quantum metric tensor under different lattice vectors can be written as $\bar{g}^{(M)}  = M\bar{g}^0M^T$ for some unimodular matrix $M$. It can be easily shown that, since $\bar{g}$ is a multiple of the identity, we always have
\begin{equation}
    \text{tr}[\bar{g}^{(M)}] \geq  \text{tr}[\bar{g}^0].
\end{equation}
This shows that the original choice of lattice vectors is optimal.

Suppose the hopping model has a reflection symmetry instead, and it can be realized if the appropriate Bravais lattice is chosen\footnote{If no additional symmetry is present, then it is the rectangular or centered rectangular lattices. If additional symmetries are present, then it may be square or hexagonal lattices, which have already been discussed. Whether the lattice should be rectangular or centered rectangular would depend on the actual model.}. Any other choice of lattice vectors is related to the reflection-symmetry-compatible choice of lattice vectors by a shear deformation. For simplicity, we can always choose our coordinate system such that the reflection symmetry of the symmetric Bravais lattice is along $x$ and $y$-directions\footnote{The hopping model may only have mirror symmetry along $x$ or $y$, but the Bravais lattice will always have both.}, and the shear matrix is given by
\begin{equation}
    M =
\begin{pmatrix}
1 & k\\
0 & 1
\end{pmatrix}.
\end{equation}

Now, let $\bar{g}^0$ be the quantum metric tensor with the symmetric choice of lattice vectors. Due to the reflection symmetry, $\bar{g}^0$ must be a diagonal matrix, so that we may write
\begin{equation}
    \bar{g}^0 = \begin{pmatrix}
\bar{g}^0_{xx} & 0\\
0 & \bar{g}^0_{yy}
\end{pmatrix}.
\end{equation}
Using $\bar{g}^{(M)}  = M\bar{g}^0M^T$ and $M$ being the shear matrix discussed above, one easily finds
\begin{equation}
    \text{tr}[\bar{g}^{(M)}] = \bar{g}^0_{xx} + (1 + k^2)\bar{g}^0_{yy}.
\end{equation}
Clearly, as $\bar{g}^0_{yy}$ is non-negative, and $k$ is a real number, we again have
\begin{equation}
    \text{tr}[\bar{g}^{(M)}] \geq  \text{tr}[\bar{g}^0].
\end{equation}
This shows that, for any choice of lattice vectors, there is some reflection symmetric choice of lattice vectors that is better (or at least as good in some degenerate cases). This shows that the optimal choice of lattice vectors must obey the reflection symmetry.

This discussion can be easily extended to glide reflection symmetry. For a crystal to have a glide reflection symmetry, the Bravais lattice must have the corresponding reflection symmetry. Just like the case of ordinary reflection symmetry, glide reflection symmetry also implies that the integral of the quantum metric tensor must be diagonal, so the argument above is equally applicable to the case of glide reflection symmetry.

Let us summarize what we have shown in this section. First, we show that, for every choice of lattice vectors, the \textit{relative} embedding that minimizes the integrated quantum metric (i.e., $I$) is the same. Second, many symmetries can only be realized if an appropriate Bravais lattice is chosen. We show that the lattice vectors that minimize $I$ (for each choice of lattice vectors, we assume optimal choice of orbital positions within the unit cell) are the ones that are compatible with the symmetry of the model.

\section{Gauge transformation}\label{app:gauge}

In Section~\ref{sec:gauge} of the main text, we have shown that a certain class of gauge transformations, given by\footnote{This includes the special case with $\bm{k}_0 = 0$.}
\begin{equation}\label{eq:linear_gauge_S}
    \hat{c}_{\bm{R},a} \rightarrow \hat{c}_{\bm{R},a}e^{i(\bm{k}_0\cdot \bm{R} + \phi_a)},
\end{equation}
does not alter quantum geometric properties, such as $I$ and $\sigma^2_{\Omega}$. The important property of this class of gauge transformations is that it does not alter the periodicity of the Hamiltonian. Conversely, it is possible to show that for a gauge transformation to preserve the periodicity of the Hamiltonian, it must be of the above given form. Gauge transformations that form the components of magnetic translation belongs to this class.

In this Appendix, we would like to consider gauge transformations that may alter the periodicity of the Hamiltonian. We shall first discuss the case of the integrated quantum metric $I$, and return to the discussion of $\sigma^2_{\Omega}$ afterwards. 

Let us consider a system that has a pair of magnetic translation symmetries $T_1$ and $T_2$ with associated translation vectors $\tilde{\bm{a}}_1$ and $\tilde{\bm{a}}_2$, such that each commute with the Hamiltonian but in general do not commute with each other. The choice of $T_1$ and $T_2$ is taken to be primitive, i.e., all other magnetic translation symmetries of the Hamiltonian can be generated from them and their inverses, and we will call the parallelogram bound by $(\tilde{\bm{a}}_1, \tilde{\bm{a}}_2)$ the \textit{primitive} unit cells. We assume that the magnetic flux is chosen in such a way that the minimal magnetic unit cell contains $q$ primitive unit cells. We would like to consider two different gauge choices, denoted with un-primed and primed quantities, with different magnetic unit cell structures, and show that their respective Hamiltonians have the same integrated quantum metric $I$.

First off, we will assume the lattice vectors of the two different magnetic unit cells are given by $\bm{a}_{1} = m \tilde{\bm{a}}_1$, $\bm{a}_{2} = n \tilde{\bm{a}}_2$ and $\bm{a}'_{1} = m' \tilde{\bm{a}}_1$, $\bm{a}'_{2} = n' \tilde{\bm{a}}_2$ respectively. The more general case of $\bm{a}_{i} = m_{i} \tilde{\bm{a}}_1 + n_{i}\tilde{\bm{a}}_2$ etc. for $i = 1,2$ can be treated similarly but is omitted here for brevity. To demonstrate that the two gauge choices give the same $I$, we adopt the following strategy: for either choice of gauge, we can artificially enlarge the unit cell to $\text{lcm}(m, m') \times \text{lcm}(n, n')$ primitive unit cells, where $\text{lcm}$ stands for ``lowest common multiple''. With this choice of supercell, the two gauges are now related by a gauge transformation of the form in Eq.~\ref{eq:linear_gauge_S}. Based on the argument in Sec.~\ref{sec:gauge} of the main text (which can be easily extended into the case of multiple bands), the two gauges, under the supercell convention, must have the same $I$. To complete the proof, we only need to show that the artificial enlargement of unit cell into the supercell does not affect $I$.

Referring to the definition of the quantum geometric tensor (whose real part is the quantum metric tensor) of a set of multiple bands given in Eq.~\ref{eq:Q_multibands}, we see that, in general, the quantum metric tensor of multiple bands is not equal to the sum of the quantum metric tensor of the constitutive bands. Nevertheless, we would like to show that, in general, formal band folding (i.e., consider an artificially enlarged unit cells without actually modifying the band structures) does not modify $I$.

To start, suppose we have a periodic system with lattice vectors $\bm{a}_1$ and $\bm{a}_2$ (the reciprocal lattice vectors are labeled as $\bm{b}_1$ and $\bm{b}_2$). One may consider an extended unit cell made of $N_1 \times N_2$ original unit cells, without altering any of the hoppings. In such a case, a single band in the original unit cell is folded into $N_1N_2$ bands in the new unit cell. Let us call the quantum metric tensor of the new unit cell $\tilde{g}_{ij}$. We would like to show that
\begin{equation}\label{eqn:g_folding}
    \tilde{g}_{ij}(\tilde{\bm{k}}) = \sum_{0 \leq n_1 < N_1}\sum_{0 \leq n_2 < N_2}g_{ij}\left(\tilde{\bm{k}} + \frac{n_1\bm{b}_1}{N_1} + \frac{n_2\bm{b}_2}{N_2}\right).
\end{equation}
That is to say, the quantum metric tensor of the folded bands is just the sum of the quantum metric tensors before folding.

We have noted that, in general, the quantum metric tensor of multiple bands is not the sum of the quantum metric tensor of individual bands. However, a sufficient condition for them to be equal is $\braket{u_m(\bm{k})|\partial_{k_i}u_n(\bm{k})} = 0$ for $m \neq n$ and $i = x,y$. The atomic basis of the extended unit cell can be written as $\ket{\bm{R},a}$, where $\bm{R} = R_1\bm{a}_1 + R_2\bm{a}_2$ for $0 \leq R_1 < N_1$ etc. is the label of the original unit cell, now reformulated as an intra-cell degree of freedom in the enlarged unit cell. From the atomic basis, we can transform into the intra-unit cell Fourier basis, where $\ket{\bm{K}, a} = \frac{1}{\sqrt{N_1N_2}}\sum_{\bm{R}}e^{i\bm{K}\cdot \bm{R}}\ket{\bm{R},a}$. One can show that, when the region in the original BZ with $\bm{k} = k_1\bm{b}_1 + k_2\bm{k}_2$ for $n_1 \leq k_1 < n_1 + 1$ and $n_2 \leq k_2 < n_2 + 1$ is folded into the mini BZ, it is folded only into the orbitals with $\bm{K} = \frac{n_1\bm{b}_1}{N_1} + \frac{n_2\bm{b}_2}{N_2}$. This means that wavefunctions of different mini bands are necessarily orthogonal to each other. This implies that $\braket{u_m(\bm{k})|\partial_{k_i}u_n(\bm{k})} = 0$ for $m \neq n$, demonstrating Eq.~\ref{eqn:g_folding}.

From this general discussion, we see that, when the $m \times n$ unit cell (and $m' \times n'$ unit cell) is formally folded into a supercell with $\text{lcm}(m, m') \times \text{lcm}(n, n')$ primitive unit cells, the integrated quantum metric is unchanged. Recalling the rest of the arguments advanced in this section, we see that different gauges with different magnetic unit cell structures give the same integrated quantum metric $I$. This means, for example, for a Hofstadter problem with $1/4$-flux per plaquette, provided the orbitals are arranged on the same grid, a $2 \times 2$ magnetic unit cell and a $4 \times 1$ magnetic unit cell yield the same $I$. Furthermore, while the $C_4$ symmetry is not manifest under the $4 \times 1$ magnetic unit cell, it is manifest (though in general still accompanied by a gauge transformation of the form given in Eq.~\ref{eq:linear_gauge_S}) under $2 \times 2$ unit cell. This implies that, to minimize $I$, the orbitals need to be arranged not just on a regular grid (which is enforced by magnetic translation symmetry), but on a \textit{square} grid, as dictated by the $C_4$ symmetry.

So far, we have focused exclusively on the integral of the trace of the quantum metric tensor $I$, without any mention of the variance of the Berry curvature $\sigma^2_{\Omega}$. For the concrete case of Hofstadter model, magnetic translation symmetries fix the real-space geometry up to an overall affine transformation (i.e., choice of lattice vectors), and we know that $\sigma^2_{\Omega}$ is independent of this affine transformation. As such, we do not need to invoke $C_4$-symmetry to determine the real-space geometry that minimizes $\sigma^2_{\Omega}$. 

Nevertheless, it remains meaningful to ask whether $\sigma^2_{\Omega}$ is invariant under gauge transformation that does not follow the simple form in Eq.~\ref{eq:linear_gauge_S}. To answer the question, we first note that the argument given for $I$ cannot be simply transplanted here, as the supercell construction would present a problem for $\sigma^2_{\Omega}$: the (abelian) Berry curvature of several bands is always the sum of Berry curvature of the constituent bands, so the following trivially holds true:
\begin{equation}\label{eqn:g_folding}
    \tilde{\Omega}(\tilde{\bm{k}}) = \sum_{0 \leq n_1 < N_1}\sum_{0 \leq n_2 < N_2}\Omega\left(\tilde{\bm{k}} + \frac{n_1\bm{b}_1}{N_1} + \frac{n_2\bm{b}_2}{N_2}\right).
\end{equation}
However, in general this does not preserve the \textit{variance} of Berry curvature, as the square of Berry curvature of the folded bands contain cross terms that have no equivalents in the case before band folding. Fortunately, as we will see, for the kind of scenarios we are interested in, the construction of supercell remains useful.

For concreteness, we will consider a system with magnetic translations $T_1$ and $T_2$ (again with associated translation vectors $\tilde{\bm{a}}_1$ and $\tilde{\bm{a}}_2$) such that they obey the commutation relation
\begin{equation}\label{eq:commutation}
    T_1 T_2 = e^{-i2\pi p/q}T_2T_1,
\end{equation}
for a pair of co-prime integers $p, q$. Suppose we have chosen the magnetic unit cell to be $m \times n$ such that $mn = q$. The Bloch states are then the eigenstates of $H$, $T_1^m$ and $T_2^n$, obeying
\begin{align}
    T^m_1\ket{\psi(\bm{k})} &= e^{im\tilde{\bm{a}}_1 \cdot \bm{k}}\ket{\psi(\bm{k})} \\
    T^n_2\ket{\psi(\bm{k})} &= e^{in\tilde{\bm{a}}_2 \cdot \bm{k}}\ket{\psi(\bm{k})},
\end{align}
and the reciprocal lattice vectors are $\frac{\tilde{\bm{b}}_1}{m}, \frac{\tilde{\bm{b}}_2}{n}$. Using the commutation relation Eq.~\ref{eq:commutation}, one finds
\begin{align}
    T^m_1(T_2\ket{\psi(\bm{k})}) &= e^{im(\tilde{\bm{a}}_1 \cdot \bm{k} - i2\pi p/q)}(T_2\ket{\psi(\bm{k})}) \\
    T^n_2(T_1\ket{\psi(\bm{k})}) &= e^{in(\tilde{\bm{a}}_2 \cdot \bm{k} + i2\pi p/q)}(T_1\ket{\psi(\bm{k})}),
\end{align}
which means
\begin{align}
    T_1 &: \bm{k} \rightarrow \bm{k} + \frac{p}{q}\tilde{\bm{b}}_2 \\
    T_2 &: \bm{k} \rightarrow \bm{k} - \frac{p}{q}\tilde{\bm{b}}_1.
\end{align}
Since $p$ and $q$ are co-prime, repeated applications of the rule imply that
\begin{equation}
    \Omega(\bm{k}) = \Omega\left(\bm{k} + \frac{\tilde{m}}{q}\tilde{\bm{b}}_1 + \frac{\tilde{n}}{q}\tilde{\bm{b}}_2\right)
\end{equation}
for arbitrary integers $\tilde{m}, \tilde{n}$. That is to say, the Berry curvature follows a periodic pattern with (momentum-space) lattice vectors $\tilde{\bm{b}}_1/q$ and $\tilde{\bm{b}}_2/q$. We note that this periodicity is independent of $m, n$, i.e., our initial choice of the shape of the magnetic unit cell.

Suppose now we have two gauge choices corresponding to magnetic unit cell with shapes $m \times n$ and $m' \times n'$ respectively, satisfying $mn = m^\prime n^\prime = q$. We want to show that the $\sigma^2_{\Omega}$ is the same for the two gauges. We have shown that, for either gauge, the Berry curvature follows periodicity given by $\tilde{\bm{b}}_1/q$ and $\tilde{\bm{b}}_2/q$, but we have not shown that within the reduced $\tilde{\bm{b}}_1/q \times \tilde{\bm{b}}_2/q$ Brillouin zone, the Berry curvature under the two different gauges are the same. To do so, we consider formally enlarging the unit cell under both gauges into $q \times q$. Using tilde to denote quantities under the supercell convention, and distinguishing the quantities arising from the two different gauges with un-primed and primed symbols, we have
\begin{align}
    \tilde{\Omega}(\bm{k}) &= q\Omega(\bm{k}) \\
    \tilde{\Omega}(\bm{k})' &= q\Omega(\bm{k})'.
\end{align}
Since, under the supercell convention, the two different gauges are related by a transformation of the form given in Eq.~\ref{eq:linear_gauge_S}, we must have
\begin{equation}
    \tilde{\Omega}(\bm{k})' = \tilde{\Omega}(\bm{k} + \bm{k}_0)
\end{equation}
for some constant $\bm{k}_0$. Therefore we also have
\begin{equation}
    \Omega(\bm{k})' = \Omega(\bm{k} + \bm{k}_0),
\end{equation}
where the range of $\bm{k}$ is unrestricted with the implicit use of the repeated zone scheme. This shows that $\sigma^2_{\Omega}$ is the same under the two different gauges.

From the above discussion, we can see that gauge transformations that alter the periodicity of the magnetic unit cell nevertheless preserve $\sigma^2_{\Omega}$, with the following caveat: let the magnetic unit cell be made of $m \times n$ primitive unit cells, in order to avoid folding ``nonequivalent'' points in the BZ together, we need $m | q$ and $n | q$\footnote{$a|b$ means $a$ is a divisor of $b$.} for the statement to hold. This clearly includes all \textit{minimal} magnetic unit cells with $mn = q$, so we have lost very little with this restriction.

\section{Ideal bands without spatial symmetry}\label{app:no_symmetry}
Denote, for some choice of gauge and magnetic unit cell, the wavefunctions of lowest Landau level as $u^{\text{LLL}}_{\bm{k}}(\bm{r})$. Then, it is known that~\cite{wang_exact_2021}
\begin{equation}
    u_{\bm{k}}(\bm{r}) = \mathcal{N}_{\bm{k}}B(\bm{r})u^{\text{LLL}}_{\bm{k}}(\bm{r}),
\end{equation}
for some $\bm{k}$-independent modulation $B(\bm{r})$ and some normalization factor $ \mathcal{N}_{\bm{k}}$, denotes an ideal band. One can also choose $B(\bm{r})$ to be supported on discrete points, such that
\begin{equation}
    u_a(\bm{k}) = \mathcal{N}_{\bm{k}}B_au^{\text{LLL}}_{\bm{k}}(\bm{r}_a),
\end{equation}
for some orbital locations $\bm{r}_a$ and weights $B_a$. Then, provided there are at least two orbitals with different locations (so that the definition is non-singular), the resulting band $u_a(\bm{k})$ is ideal. Since the choice of $\bm{r}_a$ and $B_a$ are otherwise arbitrary, one sees that, for both the continuous and discrete cases, an ideal band needs not have any spatial symmetry other than translation symmetry.

\section{Maximally localized Wannier functions}\label{app:wannier}

In this section, we discuss how our argument can be applied in the context of finding the maximally localized Wannier orbitals. We will restrict to the case where the aim is to Wannierize one isolated band, so we will drop band index throughout.

The spread of Wannier function is defined as
\begin{equation}
    \omega = \braket{r^2} - \braket{\bm{r}}^2.
\end{equation}
As is well-known, the Wannier function is dependent on the choice of gauge for Bloch wavefunctions. According to Ref.~\cite{marzari_maximally_1997}, its spread can be decomposed into two parts,
\begin{equation}
    \omega = \omega_I + \tilde{\omega},
\end{equation}
where the gauge-invariant part is given by
\begin{equation}
    \omega_I = \frac{1}{A_{\text{BZ}}}\int d\bm{k} \tr[g],
\end{equation}
where $A_{\text{BZ}}$ is the area of the Brillouin zone.  This is our familiar integrated quantum metric up to a factor. The gauge dependent part (ignoring terms that are only applicable to multiband cases) is given by
\begin{equation}\label{eq:gauge_dependent}
    \tilde{\omega} = \frac{1}{A_{\text{BZ}}}\int  d\bm{k} |\bm{A}(\bm{k}) - \bar{\bm{A}}|^2,
\end{equation}
where $\bm{A}(\bm{k})$ is the Berry connection and $\bar{\bm{A}}$ is its BZ-avearge, i.e.
\begin{equation}
    \bar{\bm{A}} =  \frac{1}{A_{\text{BZ}}}\int  d\bm{k} \bm{A}(\bm{k}).
\end{equation}

While for multiple bands, finding the maximally localized wannier functions is numerically non-trivial, for a single band, the problem can be approached analytically. In Ref.~\cite{marzari_maximally_1997}, the authors showed that Eq.~\ref{eq:gauge_dependent} is minimized by the ``Coulomb gauge'', i.e., $\nabla_{\bm{k}} \cdot \bm{A}(\bm{k}) = 0$.

Let the Berry connections in Fourier space be defined as
\begin{equation}
    \bm{A}(\bm{R}) =  \frac{1}{A_{\text{BZ}}}\int d\bm{k}e^{-i\bm{k}\cdot \bm{R}} \bm{A}(\bm{k}),
\end{equation}
with the inverse transformation given by
\begin{equation}
    \bm{A}(\bm{k}) = \sum_{\bm{R}}e^{i\bm{k}\cdot\bm{R}}\bm{A}(\bm{R}).
\end{equation}
Fourier conventions for other quantities are defined similarly. Using Fourier-space quantities, we have
\begin{equation}
    \tilde{\omega} = \sum_{\bm{R} \neq 0}|\bm{A}(\bm{R})|^2.
\end{equation}
Under the optimal gauge, $\nabla_{\bm{k}} \cdot \bm{A}(\bm{k}) = 0$, we can write
\begin{equation}\label{eq:A_chi}
    \bm{A}(\bm{k}) - \bar{\bm{A}} = \hat{z} \times \nabla_{\bm{k}}\chi(\bm{k})
\end{equation}
for some $\chi(\bm{k})$ that is periodic in BZ and whose Laplacian gives the Berry curvature, i.e.,
\begin{equation}\label{eq:laplacian}
    \nabla^2_{\bm{k}} \chi(\bm{k}) = \Omega(\bm{k}).
\end{equation}
From Eq.~\ref{eq:A_chi}, we have
\begin{equation}
    \bm{A}(\bm{R}) = i(\hat{z} \times \bm{R})\chi(\bm{R}) \quad \text{for} \ \bm{R} \neq 0,
\end{equation}
and from Eq.~\ref{eq:laplacian}, we have
\begin{equation}
    \chi(\bm{R}) = -\frac{\Omega(\bm{R})}{|\bm{R}|^2} \quad \text{for} \ \bm{R} \neq 0,
\end{equation}
where \begin{equation}
    \Omega(\bm{R}) =  \frac{1}{A_{\text{BZ}}}\int d\bm{k}e^{-i\bm{k}\cdot \bm{R}} \Omega(\bm{k}).
\end{equation}
Combining the two, we find
\begin{equation}
    \bm{A}(\bm{R})  = -i\frac{\hat{z} \times \bm{R}}{|\bm{R}|^2}\Omega(\bm{R}),
\end{equation}
so that, under the optimal gauge, we have
\begin{equation}
    \tilde{\omega} = \sum_{\bm{R} \neq 0} \frac{|\Omega(\bm{R})|^2}{|\bm{R}|^2}
\end{equation}

So far, we have not discussed embedding at all, and we have only optimized the spread of Wannier function by the choice of phases. We can now perform one further optimization by the embedding. As we have shown, the optimal spread, for a fixed embedding and optimized over Bloch phases, is 
\begin{equation}
    \omega_{\text{min}} = \frac{1}{A_{\text{BZ}}}\int d\bm{k} \tr[g] + \sum_{\bm{R} \neq 0} \frac{|\Omega(\bm{R})|^2}{|\bm{R}|^2}.
\end{equation}
The first term depends on the orbital positions as a polynomial of second order. It is easy to see that the second term should also depend on the orbital positions only up to quadratic order: we have shown $\Omega(\bm{k})$ depends on the orbital positions only up to linear order, and its Fourier transform, $\Omega(\bm{R})$ will also depend on embedding up to linear order, as Fourier transform is a linear operation. Using the same argument as we have done for $I$ and $\sigma^2_\Omega$, we see that the maximally (over gauge and orbital positions) localized Wannier orbitals must correspond to an embedding that respects the same symmetry as the hopping model.

One may object and say that the above discussion is completely trivial. After all, the spread of the Wannier orbital, defined as
\begin{equation}
    \omega = \braket{r^2} - \braket{\bm{r}}^2,
\end{equation}
is clearly a polynomial of the orbital positions up to the second order. However, the key is that, for each choice of orbital, one must optimize over the gauge under that particular embedding, and it is by no means obvious that the optimized spread remains a quadratic function of the orbital positions. Indeed, we claim that, for different orbital positions, the maximally localized Wannier functions is \textit{not} just the same Wannier function with its orbital positions changed. The proof is the following:

For a given choice of gauge, the Wannier function is defined as
\begin{equation}
    \ket{\bm{R}} =  \frac{1}{A_{\text{BZ}}}\int d\bm{k} e^{-i \bm{k} \cdot \bm{R}}\ket{\psi_{\bm{k}}}.
\end{equation}
Note that both $\ket{\bm{R}}$ and $\ket{\psi_{\bm{k}}}$ are effectively embedding independent objects (different embeddings do not change the wavefunction coefficients). Suppose we change the embedding but do not change the gauge of $\ket{\psi_{\bm{k}}}$, we have
\begin{equation}
    u_a(\bm{k}) \rightarrow e^{-i \bm{k} \cdot \bm{r}_a}u_a(\bm{k}).
\end{equation}
The Berry connection therefore transforms as
\begin{equation}
    \bm{A}(\bm{k}) \rightarrow \bm{A}(\bm{k}) + \sum_a \bm{r}_a |u_a(\bm{k})|^2,
\end{equation}
and we see that the optimal gauge condition (for a single band),
\begin{equation}
    \nabla_{\bm{k}} \cdot \bm{A}(\bm{k}) = 0,
\end{equation}
is \textit{not} invariant under the above change.

\section{Global minimum of quantum volume}\label{app:quantum_volume}

Consider a triangular lattice with bond vectors $\bm{a}_1 = (1, 0)$, $\bm{a}_2 = (-\frac{1}{2}, \frac{\sqrt{3}}{2})$, $\bm{a}_3 = (-\frac{1}{2}, -\frac{\sqrt{3}}{2})$, and define $k_a = \bm{k} \cdot \bm{a}_a$ for $a = 1,2,3$. On this triangular lattice, we construct the following  3-band model:

\begin{equation}\label{eq:model}
u_a(\bm{k})=\tfrac{1}{\sqrt3}\,e^{i\lambda\cos k_a},\
H(\bm{k})=-\ket{u(\bm{k})}\bra{u(\bm{k})},\ \lambda\neq0 .
\end{equation}
The wavefunctions as written satisfy $u_a(\bm{k} + \bm{G}) = u_a(\bm{k})$ for any reciprocal lattice vector $\bm{G}$. This means that we have assumed the trivial embedding (i.e., placing all three orbitals at the same lattice site) when writing the wavefunctions.

It can be readily verified that the model satisfies 3-fold rotation symmetry $C_3$ with a permutation of the orbital basis, namely
\begin{equation}
    \sigma_{C_3}(\bm{R}, a) = (C_3\bm{R}, a +1),
\end{equation}
and inversion symmetry $P$
\begin{equation}
    \sigma_P(\bm{R}, a) = (-\bm{R}, a).
\end{equation}
We note that the inversion symmetry is only preserved, in the sense defined in Sec.~\ref{subsec:setup}, when all orbitals are embedded at the same location. This is true despite there are multiple inversion-symmetric positions within a unit cell, because, for example, if one places each orbital into a distinct $M$-point in the unit cell, it cannot be true that $P$ maps $\bm{R}$ to $-\bm{R}$ for all three orbitals without orbital dependent shift. As such, the only fully symmetric embedding is the trivial embedding. 

\begin{figure}[t]
\centering
\includegraphics[width=\linewidth]{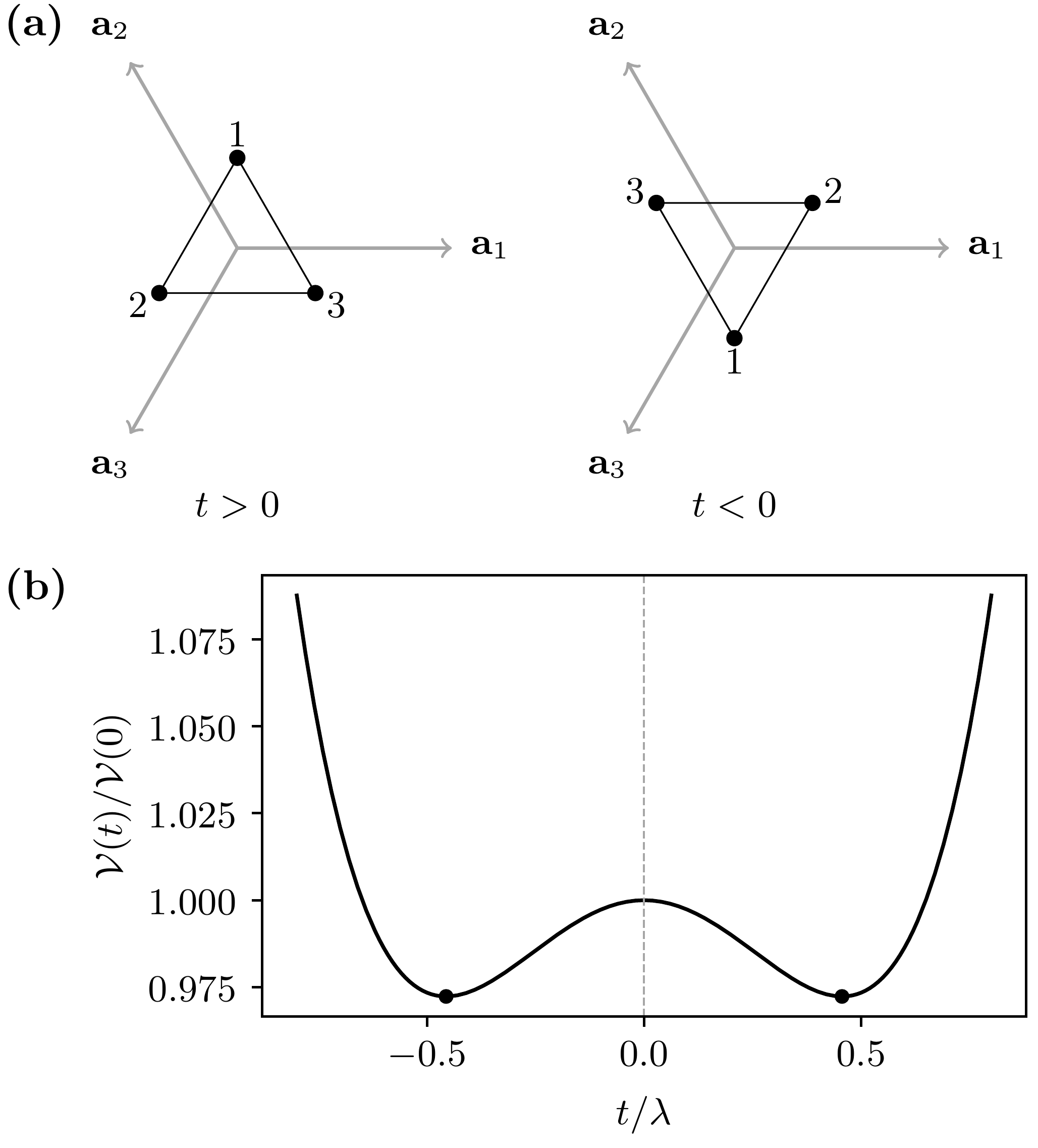}
\caption{(a) $C_3$-symmetric embeddings: orbital $a$ is displaced by $(t/\sqrt3)\,\hat z\times\bm{a}_a$, forming an equilateral triangle of side $|t|$. (b) $\mathcal{V}(t)/\mathcal{V}(0)$ along the one-parameter family of $C_3$-symmetric embedding. Dots mark $\pm t_*$.}
\label{fig:main}
\end{figure}

Allowing inversion symmetry $P$ to be broken while preserving $C_3$, we define the following family of embedding parameterized by a single parameter $t$:
\begin{equation}
\bm{r}_a(t)=\tfrac{t}{\sqrt3}\,\hat z\times\bm{a}_a,
\end{equation}
as illustrated in Fig.~\ref{fig:main}. $t = 0$ is the only inversion symmetric parameter in this family of embedding. The quantum volume of the lowest band, as a function of $t$, is given by
\begin{equation}\label{eq:Vt}
\mathcal{V}(t)=\frac{\sqrt3}{9}\int_{[0,2\pi)^2}\big|\lambda^2J(\bm{k})+t^2\big|\,dk_1dk_2,
\end{equation}
where
\begin{equation}
J=\sin k_1\sin k_2+\sin k_2\sin k_3+\sin k_3\sin k_1,
\end{equation}
with $k_3 = - k_1 - k_2$. 

In Fig.~\ref{fig:main}(b), we show $\mathcal{V}(t)/\mathcal{V}(0)$ as a function of $t/\lambda$, and we observe that the quantum volume is minimized within this one-parameter family by $t = \pm t_*$ for some $t_* \neq 0$. Since $t = 0$ is the only fully symmetric embedding up to an overall shift but is \emph{not} the global minimizer (in this case, not even a local minimizer) of the quantum volume, this shows that the global minimum of the quantum volume does not always correspond
to a symmetric embedding.

\bibliography{references}

@article{claassen_position-momentum_2015,
  title = {Position-{{Momentum Duality}} and {{Fractional Quantum Hall Effect}} in {{Chern Insulators}}},
  author = {Claassen, Martin and Lee, Ching Hua and Thomale, Ronny and Qi, Xiao-Liang and Devereaux, Thomas P.},
  year = {2015},
  month = jun,
  journal = {Physical Review Letters},
  volume = {114},
  number = {23},
  pages = {236802},
  publisher = {American Physical Society},
  doi = {10.1103/PhysRevLett.114.236802},
  urldate = {2024-02-27}
}

@article{dobardzic_geometrical_2013,
  title = {Geometrical Description of Fractional {{Chern}} Insulators Based on Static Structure Factor Calculations},
  author = {Dobard{\v z}i{\'c}, E. and Milovanovi{\'c}, M. V. and Regnault, N.},
  year = {2013},
  month = sep,
  journal = {Physical Review B},
  volume = {88},
  number = {11},
  pages = {115117},
  publisher = {American Physical Society},
  doi = {10.1103/PhysRevB.88.115117},
  urldate = {2023-02-27}
}

@article{herzog-arbeitmanManybodySuperconductivityTopological2026,
  title = {Many-Body Superconductivity in Topological Flat Bands},
  author = {{Herzog-Arbeitman}, Jonah and Chew, Aaron and Huhtinen, Kukka-Emilia and T{\"o}rm{\"a}, P{\"a}ivi and Bernevig, B. Andrei},
  year = 2026,
  month = jul,
  journal = {Communications Physics},
  publisher = {Nature Publishing Group},
  issn = {2399-3650},
  doi = {10.1038/s42005-026-02732-2},
  urldate = {2026-07-28},
  copyright = {2026 The Author(s)},
  langid = {english}
}

@article{jackson_geometric_2015,
  title = {Geometric Stability of Topological Lattice Phases},
  author = {Jackson, T. S. and M{\"o}ller, Gunnar and Roy, Rahul},
  year = {2015},
  month = nov,
  journal = {Nature Communications},
  volume = {6},
  number = {1},
  pages = {8629},
  publisher = {Nature Publishing Group},
  issn = {2041-1723},
  doi = {10.1038/ncomms9629},
  urldate = {2022-10-24},
  copyright = {2015 The Author(s)},
  langid = {english}
}

@article{ledwith_family_2022,
  title = {Family of {{Ideal Chern Flatbands}} with {{Arbitrary Chern Number}} in {{Chiral Twisted Graphene Multilayers}}},
  author = {Ledwith, Patrick J. and Vishwanath, Ashvin and Khalaf, Eslam},
  year = {2022},
  month = apr,
  journal = {Physical Review Letters},
  volume = {128},
  number = {17},
  pages = {176404},
  publisher = {American Physical Society},
  doi = {10.1103/PhysRevLett.128.176404},
  urldate = {2022-11-10}
}

@article{lee_band_2017,
  title = {Band Structure Engineering of Ideal Fractional {{Chern}} Insulators},
  author = {Lee, Ching Hua and Claassen, Martin and Thomale, Ronny},
  year = {2017},
  month = oct,
  journal = {Physical Review B},
  volume = {96},
  number = {16},
  pages = {165150},
  publisher = {American Physical Society},
  doi = {10.1103/PhysRevB.96.165150},
  urldate = {2022-11-21}
}

@article{ozawa_relations_2021,
  title = {Relations between Topology and the Quantum Metric for {{Chern}} Insulators},
  author = {Ozawa, Tomoki and Mera, Bruno},
  year = {2021},
  month = jul,
  journal = {Physical Review B},
  volume = {104},
  number = {4},
  pages = {045103},
  publisher = {American Physical Society},
  doi = {10.1103/PhysRevB.104.045103},
  urldate = {2022-10-19}
}

@article{parameswaran_fractional_2012,
  title = {Fractional {{Chern}} Insulators and the {$W_\infty$} Algebra},
  author = {Parameswaran, S. A. and Roy, R. and Sondhi, S. L.},
  year = {2012},
  month = jun,
  journal = {Physical Review B},
  volume = {85},
  number = {24},
  pages = {241308},
  publisher = {American Physical Society},
  doi = {10.1103/PhysRevB.85.241308},
  urldate = {2022-10-14}
}

@article{roy_band_2014,
  title = {Band Geometry of Fractional Topological Insulators},
  author = {Roy, Rahul},
  year = {2014},
  month = oct,
  journal = {Physical Review B},
  volume = {90},
  number = {16},
  pages = {165139},
  publisher = {American Physical Society},
  doi = {10.1103/PhysRevB.90.165139},
  urldate = {2022-10-14}
}

@article{simon_contrasting_2020,
  title = {Contrasting Lattice Geometry Dependent versus Independent Quantities: {{Ramifications}} for {{Berry}} Curvature, Energy Gaps, and Dynamics},
  shorttitle = {Contrasting Lattice Geometry Dependent versus Independent Quantities},
  author = {Simon, Steven H. and Rudner, Mark S.},
  year = {2020},
  month = oct,
  journal = {Physical Review B},
  volume = {102},
  number = {16},
  pages = {165148},
  publisher = {American Physical Society},
  doi = {10.1103/PhysRevB.102.165148},
  urldate = {2022-10-21}
}

@article{wang_exact_2021,
  title = {Exact {{Landau Level Description}} of {{Geometry}} and {{Interaction}} in a {{Flatband}}},
  author = {Wang, Jie and Cano, Jennifer and Millis, Andrew J. and Liu, Zhao and Yang, Bo},
  year = {2021},
  month = dec,
  journal = {Physical Review Letters},
  volume = {127},
  number = {24},
  pages = {246403},
  publisher = {American Physical Society},
  doi = {10.1103/PhysRevLett.127.246403},
  urldate = {2022-11-21}
}

@article{wang_hierarchy_2022,
  title = {Hierarchy of {{Ideal Flatbands}} in {{Chiral Twisted Multilayer Graphene Models}}},
  author = {Wang, Jie and Liu, Zhao},
  year = {2022},
  month = apr,
  journal = {Physical Review Letters},
  volume = {128},
  number = {17},
  pages = {176403},
  publisher = {American Physical Society},
  doi = {10.1103/PhysRevLett.128.176403},
  urldate = {2023-01-04}
}

@article{wang_origin_2023,
  title = {Origin of Model Fractional {{Chern}} Insulators in All Topological Ideal Flatbands: {{Explicit}} Color-Entangled Wave Function and Exact Density Algebra},
  shorttitle = {Origin of Model Fractional {{Chern}} Insulators in All Topological Ideal Flatbands},
  author = {Wang, Jie and Klevtsov, Semyon and Liu, Zhao},
  year = {2023},
  month = jun,
  journal = {Physical Review Research},
  volume = {5},
  number = {2},
  pages = {023167},
  publisher = {American Physical Society},
  doi = {10.1103/PhysRevResearch.5.023167},
  urldate = {2024-02-26}
}

@article{wu_zoology_2012,
  title = {Zoology of Fractional {{Chern}} Insulators},
  author = {Wu, Yang-Le and Bernevig, B. Andrei and Regnault, N.},
  year = {2012},
  month = feb,
  journal = {Physical Review B},
  volume = {85},
  number = {7},
  pages = {075116},
  publisher = {American Physical Society},
  doi = {10.1103/PhysRevB.85.075116},
  urldate = {2023-02-27}
}

@article{Bauer_2016,
  title = {Quantum geometry and stability of the fractional quantum Hall effect in the Hofstadter model},
  author = {Bauer, David and Jackson, T. S. and Roy, Rahul},
  journal = {Phys. Rev. B},
  volume = {93},
  issue = {23},
  pages = {235133},
  numpages = {11},
  year = {2016},
  month = {Jun},
  publisher = {American Physical Society},
  doi = {10.1103/PhysRevB.93.235133},
  url = {https://link.aps.org/doi/10.1103/PhysRevB.93.235133}
}

@article{shankar_murthy,
  title = {Hamiltonian theory of fractionally filled Chern bands},
  author = {Murthy, Ganpathy and Shankar, R.},
  journal = {Phys. Rev. B},
  volume = {86},
  issue = {19},
  pages = {195146},
  numpages = {15},
  year = {2012},
  month = {Nov},
  publisher = {American Physical Society},
  doi = {10.1103/PhysRevB.86.195146},
  url = {https://link.aps.org/doi/10.1103/PhysRevB.86.195146}
}

@article{liu_theory_2025,
  title = {Theory of {{Generalized Landau Levels}} and {{Its Implications}} for {{Non-Abelian States}}},
  author = {Liu, Zhao and Mera, Bruno and Fujimoto, Manato and Ozawa, Tomoki and Wang, Jie},
  year = 2025,
  month = jul,
  journal = {Physical Review X},
  volume = {15},
  number = {3},
  pages = {031019},
  publisher = {American Physical Society},
  doi = {10.1103/1zg9-qbd6},
  urldate = {2026-07-13}
}

@article{mera_kahler_2021,
  title = {K{\"a}hler Geometry and {{Chern}} Insulators: {{Relations}} between Topology and the Quantum Metric},
  shorttitle = {K{\textbackslash}"ahler Geometry and {{Chern}} Insulators},
  author = {Mera, Bruno and Ozawa, Tomoki},
  year = {2021},
  month = jul,
  journal = {Physical Review B},
  volume = {104},
  number = {4},
  pages = {045104},
  publisher = {American Physical Society},
  doi = {10.1103/PhysRevB.104.045104},
  urldate = {2024-09-30}
}

@article{mera_engineering_2021,
  title = {Engineering Geometrically Flat {{Chern}} Bands with {{Fubini-Study K}}{\"a}hler Structure},
  author = {Mera, Bruno and Ozawa, Tomoki},
  year = {2021},
  month = sep,
  journal = {Physical Review B},
  volume = {104},
  number = {11},
  pages = {115160},
  publisher = {American Physical Society},
  doi = {10.1103/PhysRevB.104.115160},
  urldate = {2024-09-30}
}

@article{kapit_exact_2010,
  title = {Exact {{Parent Hamiltonian}} for the {{Quantum Hall States}} in a {{Lattice}}},
  author = {Kapit, Eliot and Mueller, Erich},
  year = {2010},
  month = nov,
  journal = {Physical Review Letters},
  volume = {105},
  number = {21},
  pages = {215303},
  publisher = {American Physical Society},
  doi = {10.1103/PhysRevLett.105.215303},
  urldate = {2022-10-19}
}

@article{goerbig_fractional_2012,
  title = {From Fractional {{Chern}} Insulators to a Fractional Quantum Spin Hall Effect},
  author = {Goerbig, M. O.},
  year = {2012},
  month = jan,
  journal = {The European Physical Journal B},
  volume = {85},
  number = {1},
  pages = {15},
  issn = {1434-6036},
  doi = {10.1140/epjb/e2011-20857-6},
  urldate = {2024-09-30},
  langid = {english}
}

@article{ledwith_vortexability_2023,
  title = {Vortexability: {{A}} Unifying Criterion for Ideal Fractional {{Chern}} Insulators},
  shorttitle = {Vortexability},
  author = {Ledwith, Patrick J. and Vishwanath, Ashvin and Parker, Daniel E.},
  year = 2023,
  month = nov,
  journal = {Physical Review B},
  volume = {108},
  number = {20},
  pages = {205144},
  publisher = {American Physical Society},
  doi = {10.1103/PhysRevB.108.205144},
  urldate = {2026-04-12}
}

@article{fujimoto_higher_2025,
  title = {Higher {{Vortexability}}: {{Zero-Field Realization}} of {{Higher Landau Levels}}},
  shorttitle = {Higher {{Vortexability}}},
  author = {Fujimoto, Manato and Parker, Daniel E. and Dong, Junkai and Khalaf, Eslam and Vishwanath, Ashvin and Ledwith, Patrick},
  year = 2025,
  month = mar,
  journal = {Physical Review Letters},
  volume = {134},
  number = {10},
  pages = {106502},
  publisher = {American Physical Society},
  doi = {10.1103/PhysRevLett.134.106502},
  urldate = {2026-07-13}
}

@article{mera_uniqueness_2024,
  title = {Uniqueness of {{Landau}} Levels and Their Analogs with Higher {{Chern}} Numbers},
  author = {Mera, Bruno and Ozawa, Tomoki},
  year = {2024},
  month = sep,
  journal = {Physical Review Research},
  volume = {6},
  number = {3},
  pages = {033238},
  publisher = {American Physical Society},
  doi = {10.1103/PhysRevResearch.6.033238},
  urldate = {2024-10-09}
}

@article{liang_band_2017,
  title = {Band Geometry, {{Berry}} Curvature, and Superfluid Weight},
  author = {Liang, Long and Vanhala, Tuomas I. and Peotta, Sebastiano and Siro, Topi and Harju, Ari and T{\"o}rm{\"a}, P{\"a}ivi},
  year = 2017,
  month = jan,
  journal = {Physical Review B},
  volume = {95},
  number = {2},
  pages = {024515},
  publisher = {American Physical Society},
  doi = {10.1103/PhysRevB.95.024515},
  urldate = {2026-07-14}
}

@article{qi_topological_2006,
  title = {Topological Quantization of the Spin {{Hall}} Effect in Two-Dimensional Paramagnetic Semiconductors},
  author = {Qi, Xiao-Liang and Wu, Yong-Shi and Zhang, Shou-Cheng},
  year = 2006,
  month = aug,
  journal = {Physical Review B},
  volume = {74},
  number = {8},
  pages = {085308},
  publisher = {American Physical Society},
  doi = {10.1103/PhysRevB.74.085308},
  urldate = {2024-06-03}
}

@article{marzari_maximally_1997,
  title = {Maximally Localized Generalized {{Wannier}} Functions for Composite Energy Bands},
  author = {Marzari, Nicola and Vanderbilt, David},
  year = 1997,
  month = nov,
  journal = {Physical Review B},
  volume = {56},
  number = {20},
  pages = {12847--12865},
  publisher = {American Physical Society},
  doi = {10.1103/PhysRevB.56.12847},
  urldate = {2026-06-01}
}

@article{lim_geometry_2015,
  author  = {Lim, Lih-King and Fuchs, Jean-No{\"e}l and Montambaux, Gilles},
  title   = {Geometry of Bloch States Probed by St{\"u}ckelberg Interferometry},
  journal = {Physical Review A},
  volume  = {92},
  pages   = {063627},
  year    = {2015},
  doi     = {10.1103/PhysRevA.92.063627}
}

@article{cooper_topological_2019,
  author  = {Cooper, N. R. and Dalibard, Jean and Spielman, I. B.},
  title   = {Topological Bands for Ultracold Atoms},
  journal = {Reviews of Modern Physics},
  volume  = {91},
  pages   = {015005},
  year    = {2019},
  doi     = {10.1103/RevModPhys.91.015005}
}

@book{vanderbilt_berry_2018,
  title = {Berry {{Phases}} in {{Electronic Structure Theory}}: {{Electric Polarization}}, {{Orbital Magnetization}} and {{Topological Insulators}}},
  shorttitle = {Berry {{Phases}} in {{Electronic Structure Theory}}},
  author = {Vanderbilt, David},
  year = 2018,
  publisher = {Cambridge University Press},
  address = {Cambridge},
  doi = {10.1017/9781316662205},
  urldate = {2026-07-16},
  isbn = {978-1-107-15765-1}
}

@article{julku_geometric_2016,
  title = {Geometric {{Origin}} of {{Superfluidity}} in the {{Lieb-Lattice Flat Band}}},
  author = {Julku, Aleksi and Peotta, Sebastiano and Vanhala, Tuomas I. and Kim, Dong-Hee and T{\"o}rm{\"a}, P{\"a}ivi},
  year = 2016,
  month = jul,
  journal = {Physical Review Letters},
  volume = {117},
  number = {4},
  pages = {045303},
  publisher = {American Physical Society},
  doi = {10.1103/PhysRevLett.117.045303},
  urldate = {2026-07-14}
}

@article{peotta_superfluidity_2015,
  title = {Superfluidity in Topologically Nontrivial Flat Bands},
  author = {Peotta, Sebastiano and T{\"o}rm{\"a}, P{\"a}ivi},
  year = 2015,
  month = nov,
  journal = {Nature Communications},
  volume = {6},
  number = {1},
  pages = {8944},
  publisher = {Nature Publishing Group},
  issn = {2041-1723},
  doi = {10.1038/ncomms9944},
  urldate = {2025-10-03},
  copyright = {2015 The Author(s)},
  langid = {english}
}

@article{tam_geometry-independent_2024,
  title = {Geometry-Independent Superfluid Weight in Multiorbital Lattices from the Generalized Random Phase Approximation},
  author = {Tam, Minh and Peotta, Sebastiano},
  year = 2024,
  month = mar,
  journal = {Physical Review Research},
  volume = {6},
  number = {1},
  pages = {013256},
  publisher = {American Physical Society},
  doi = {10.1103/PhysRevResearch.6.013256},
  urldate = {2026-07-13}
}

@article{herzog-arbeitman_superfluid_2022,
  title = {Superfluid {{Weight Bounds}} from {{Symmetry}} and {{Quantum Geometry}} in {{Flat Bands}}},
  author = {{Herzog-Arbeitman}, Jonah and Peri, Valerio and Schindler, Frank and Huber, Sebastian D. and Bernevig, B. Andrei},
  year = 2022,
  month = feb,
  journal = {Physical Review Letters},
  volume = {128},
  number = {8},
  pages = {087002},
  publisher = {American Physical Society},
  doi = {10.1103/PhysRevLett.128.087002},
  urldate = {2024-05-21}
}

@article{haldane_model_1988,
  title = {Model for a {{Quantum Hall Effect}} without {{Landau Levels}}: {{Condensed-Matter Realization}} of the "{{Parity Anomaly}}"},
  shorttitle = {Model for a {{Quantum Hall Effect}} without {{Landau Levels}}},
  author = {Haldane, F. D. M.},
  year = 1988,
  month = oct,
  journal = {Physical Review Letters},
  volume = {61},
  number = {18},
  pages = {2015--2018},
  publisher = {American Physical Society},
  doi = {10.1103/PhysRevLett.61.2015},
  urldate = {2022-10-19}
}

@article{huhtinen_revisiting_2022,
  title = {Revisiting Flat Band Superconductivity: {{Dependence}} on Minimal Quantum Metric and Band Touchings},
  shorttitle = {Revisiting Flat Band Superconductivity},
  author = {Huhtinen, Kukka-Emilia and {Herzog-Arbeitman}, Jonah and Chew, Aaron and Bernevig, Bogdan A. and T{\"o}rm{\"a}, P{\"a}ivi},
  year = 2022,
  month = jul,
  journal = {Physical Review B},
  volume = {106},
  number = {1},
  pages = {014518},
  publisher = {American Physical Society},
  doi = {10.1103/PhysRevB.106.014518},
  urldate = {2025-10-03}
}

@article{fuchs_orbital_2021,
  title = {Orbital Embedding and Topology of One-Dimensional Two-Band Insulators},
  author = {Fuchs, Jean-No{\"e}l and Pi{\'e}chon, Fr{\'e}d{\'e}ric},
  year = 2021,
  month = dec,
  journal = {Physical Review B},
  volume = {104},
  number = {23},
  pages = {235428},
  publisher = {American Physical Society},
  doi = {10.1103/PhysRevB.104.235428},
  urldate = {2026-07-13}
}

@article{fruchart_parallel_2014,
  title = {Parallel Transport and Band Theory in Crystals},
  author = {Fruchart, Michel and Carpentier, David and Gaw{\k e}dzki, Krzysztof},
  year = 2014,
  month = jun,
  journal = {Europhysics Letters},
  volume = {106},
  number = {6},
  pages = {60002},
  publisher = {{EDP Sciences, IOP Publishing and Societ\`a Italiana di Fisica}},
  issn = {0295-5075},
  doi = {10.1209/0295-5075/106/60002},
  urldate = {2026-07-13},
  langid = {english}
}

@article{dobardzic_effective_2014,
  title = {Effective Description of {{Chern}} Insulators},
  author = {Dobard{\v z}i{\'c}, E. and Dimitrijevi{\'c}, M. and Milovanovi{\'c}, M. V.},
  year = 2014,
  month = jun,
  journal = {Physical Review B},
  volume = {89},
  number = {23},
  pages = {235424},
  publisher = {American Physical Society},
  doi = {10.1103/PhysRevB.89.235424},
  urldate = {2026-07-13}
}

@article{thouless_quantized_1982,
  title = {Quantized {{Hall Conductance}} in a {{Two-Dimensional Periodic Potential}}},
  author = {Thouless, D. J. and Kohmoto, M. and Nightingale, M. P. and {den Nijs}, M.},
  year = 1982,
  month = aug,
  journal = {Physical Review Letters},
  volume = {49},
  number = {6},
  pages = {405--408},
  publisher = {American Physical Society},
  doi = {10.1103/PhysRevLett.49.405},
  urldate = {2022-10-19}
}

@article{xiao_berry_2010,
  title = {Berry Phase Effects on Electronic Properties},
  author = {Xiao, Di and Chang, Ming-Che and Niu, Qian},
  year = 2010,
  month = jul,
  journal = {Reviews of Modern Physics},
  volume = {82},
  number = {3},
  pages = {1959--2007},
  publisher = {American Physical Society},
  doi = {10.1103/RevModPhys.82.1959},
  urldate = {2024-10-02}
}

@article{shi_effects_2026,
  title = {Effects of {{Berry}} Curvature on Ideal Fractional {{Chern}} Insulator Many-Body Gaps},
  author = {Shi, Jingtian and Cano, Jennifer and {Morales-Dur{\'a}n}, Nicol{\'a}s},
  year = 2026,
  month = jun,
  journal = {Physical Review Research},
  volume = {8},
  number = {2},
  pages = {L022045},
  publisher = {American Physical Society},
  doi = {10.1103/sfm3-f1kp},
  urldate = {2026-07-13}
}

@misc{li_quantum-geometric_2026,
  title = {Quantum-{{Geometric Design}} of {{Lattice Generalized Landau Levels}}},
  author = {Li, Bohao and Wu, Fengcheng},
  year = 2026,
  month = jul,
  number = {arXiv:2607.08702},
  eprint = {2607.08702},
  primaryclass = {cond-mat.mes-hall},
  publisher = {arXiv},
  doi = {10.48550/arXiv.2607.08702},
  urldate = {2026-07-21},
  archiveprefix = {arXiv}
}

@misc{to_appear,
title = {Ideal Bands in Tight-Binding Models},
author = {Zhao, Lexu and Ge, Yang and Ryu, Shinsei and Yu, Jiabin},
year = 2026,
note = {to appear}
}

\end{document}